\documentclass[reprint,pre,amsmath,amssymb,floatfix,showpacs]{revtex4-1}
\usepackage{graphicx}
\usepackage{bm} 
\usepackage{threeparttable} 
\usepackage{color}

\newcommand{\ivec}{{\bf i}}
\newcommand{\Ivec}{{\bf I}}
\newcommand{\numat}{{\underline {\nu}}}
\newcommand{\mumat}{{\underline {\mu}}}
\newcommand{\Pmat}{{\underline {\rm P}}}
\newcommand{\Smat}{{\underline {\rm S}}}

\newcommand{\Ptilde}{{\underline {\tilde {\rm P}}}}
\newcommand{\Stilde}{{\underline {\tilde {\rm S}}}}
\newcommand{\Phimat}{{\underline {\Phi}}}
\newcommand{\Thetamat}{{\underline {\Theta}}}
\newcommand{\Amat}{{\underline {\rm A}}}
\newcommand{\AmatT}{{\underline {\rm A}}^{\rm T}}
\newcommand{\Bmat}{{\underline {\rm B}}}
\newcommand{\BmatT}{{\underline {\rm B}}^{\rm T}}
\newcommand{\Umat}{{\underline {\rm U}}}
\newcommand{\UmatT}{{\underline {\rm U}}^{\rm T}}
\newcommand{\Rmat}{{\underline {\rm R}}}
\newcommand{\RmatT}{{\underline {\rm R}}^{\rm T}}
\newcommand{\MV}{}
\newcommand{\MM}{\,}

\begin{document}

\title{Framework for constructing generic Jastrow correlation factors}

\author{P.~\surname{L\'opez R\'{\i}os}}
\affiliation{Theory of Condensed Matter Group, Cavendish Laboratory,
University of Cambridge, J.~J.~Thomson Avenue, Cambridge CB3 0HE,
United Kingdom}

\author{P.~Seth}
\affiliation{Theory of Condensed Matter Group, Cavendish Laboratory,
University of Cambridge, J.~J.~Thomson Avenue, Cambridge CB3 0HE,
United Kingdom}

\author{N.~D.~Drummond}
\affiliation{Department of Physics, Lancaster University,
Lancaster LA1 4YB, United Kingdom}

\author{R.~J.~Needs}
\affiliation{Theory of Condensed Matter Group, Cavendish Laboratory,
University of Cambridge, J.~J.~Thomson Avenue, Cambridge CB3 0HE,
United Kingdom}

\date{\today}

\begin{abstract}
  We have developed a flexible framework for constructing Jastrow
  factors which allows for the introduction of terms involving
  arbitrary numbers of particles.
  The use of various three- and four-body Jastrow terms in quantum
  Monte Carlo calculations is investigated, including a four-body van
  der Waals-like term, and anisotropic terms.
  We have tested these Jastrow factors on one- and two-dimensional
  homogeneous electron gases, the Be, B, and O atoms, and the BeH,
  H$_2$O, N$_2$, and H$_2$ molecules.
  Our optimized Jastrow factors retrieve more than 90\% of the fixed-node
  diffusion Monte Carlo
  correlation energy in variational Monte Carlo for each system studied.
\end{abstract}

\pacs{02.70.Ss, 31.25.-v, 71.10.-w, 71.15.-m}

\maketitle

\section{Introduction}
\label{sec:intro}

The variational and diffusion quantum Monte Carlo methods (VMC and
DMC) are zero-temperature stochastic techniques for evaluating the
expectation values of time-independent operators
\cite{foulkes_quantum_2001,kolorenc_applications_2011,austin_quantum_2011}.
These methods are particularly well suited for calculating the
ground-state energies of large assemblies of interacting quantum
particles.
The central object is an approximate trial wave function whose
accuracy controls the computed energy and the intrinsic statistical
fluctuations in the calculations.
It is therefore of particular importance to develop accurate trial
wave functions.

Expectation values in VMC are evaluated using importance-sampled Monte
Carlo integration.
In DMC the ground state is projected out by evolving the Schr\"odinger
equation in imaginary time.
Such projector methods suffer from a fermion sign problem, in which
errors in the propagation increase exponentially in imaginary time as
the algorithm amplifies any spurious component of the lower-energy
bosonic state.
This problem may be evaded in DMC by employing the fixed-node
approximation \cite{anderson_fixed_node_1976}, in which the nodal
surface is fixed to be that of a suitable trial wave function.
The resulting DMC energy is greater than or equal to the exact energy
and less than or equal to the VMC energy computed with the same trial
wave function.
The VMC energy depends on the entire trial wave function, but the DMC
energy depends only on the nodal surface of the trial wave function.

One of the appealing features of VMC and DMC is that virtually any
form of trial wave function can be used.
The main criteria are that the wave function must obey the correct
symmetry under particle exchange, it should be flexible enough to
describe the system of interest, and that it should be possible to
evaluate it rapidly.
The analytic properties and normalizability of the trial wave function
must be such that the energy expectation value is well-defined.
The simplest fermionic wave function is a Slater determinant, which
describes exchange but not correlation.
Multideterminant wave functions, pairing wave functions such as
geminals \cite{casula_correlated_2004}, and backflow transformations
\cite{lopez_rios_inhomogeneous_2006} can also be used.
The most fruitful method of going beyond the Slater determinant is,
however, to multiply it by a Jastrow factor
\cite{foulkes_quantum_2001}, which leads to the Slater-Jastrow wave
function.
The Jastrow factor is normally chosen to depend on the interparticle
separations, which introduces correlation into the wave function.
The introduction of a Jastrow factor often leads to the recovery of
80\% or more of the correlation energy of electronic systems
\cite{drummond_jastrow_2004}.

The Jastrow factor is chosen to be everywhere positive and symmetric
with respect to the exchange of identical particles in order to
maintain the nodal surface defined by the rest of the wave function.
One of the features of the Jastrow factor is that it can conveniently
be used to enforce the Kato cusp conditions
\cite{kato_eigenfunctions_1957}, which determine the behavior of the
wave function when two charged particles approach one another.
Enforcing the Kato cusp conditions does not necessarily improve the
variational energy, but the reduction in the statistical fluctuations
in the energy is often very important.

DMC can be viewed as VMC with a perfect Jastrow factor, but improving
the Jastrow factor can improve DMC calculations in several ways.
The DMC algorithm is subject to time-step errors and to (normally very
small) population-control errors \cite{umrigar_diffusion_1993} that
are reduced by improving the trial wave function.
Evaluating expectation values of operators that do not commute with
the Hamiltonian is not straightforward in DMC, but using highly
accurate trial wave functions helps in achieving more accurate
results.
Similar considerations apply when using nonlocal pseudopotentials,
which involves making approximations that are ameliorated by improving
the trial wave function \cite{mitas_nonlocal_1991,casula_size-consistent_2010}.
As the fundamental limitation on the accuracy of DMC is the quality of
the nodal surface, it is desirable to use trial wave functions with
optimizable nodal surfaces as afforded by, for example,
multideterminant wave functions and backflow transformations.
A good Jastrow factor can account for the bulk of the dynamical
correlation energy, which allows the optimization of parameters that
affect the nodal surface to achieve a better nodal surface.

Here we introduce a highly flexible form of Jastrow factor which
allows for the introduction of a variety of terms involving arbitrary
numbers of particles.
Our main motivation is to be able to implement quickly different
functional forms and explore the importance of different correlations
in any physical system we study.

Jastrow factors correlating several electrons have been used in
earlier calculations, such as those of
Refs.~\cite{huang_accuracy_1997,kwon_effects_1993,alexander_ground_2004}.
We study the effects of various three-body Jastrow terms and
introduce a four-body van der Waals-like term.
We also construct anisotropic Jastrow factors that can capture the
natural symmetries of a system.
We have successfully applied these Jastrow factors to a variety of
systems, and we report results for the one- and two-dimensional
homogeneous electron gases, the Be, B, and O atoms, and the BeH,
N$_2$, H$_2$O, and H$_2$ molecules.
Our VMC and DMC calculations were performed using the \textsc{casino}
package \cite{needs_continuum_2010}.
Hartree atomic units are used throughout
($\hbar=|e|=m_e=4\pi\epsilon_0=1$).
The structure of the paper is as follows.
We describe the form and properties of the general Jastrow factor in
Sec.~\ref{sec:gjastrow}.
Specific examples of the construction of Jastrow terms are given
in Sec.~\ref{sec:construction}, and results obtained using them
are presented in Sec.~\ref{sec:results}.
Finally, we draw our conclusions in Sec.~\ref{sec:conclusions}.
Implementation details are given in Appendix~\ref{app:implementation},
and an example can be found in Appendix~\ref{app:example}.
We report only summaries of our data in this paper
\cite{supplemental}.

\section{Construction of a generic Jastrow factor}
\label{sec:gjastrow}

Quantum Monte Carlo (QMC) methods can be applied to systems which can
be generically described as an ensemble of $N$ quantum particles and
$M$ sources of external potential.
The most common type of QMC simulations are electronic calculations,
where the quantum particles are electrons and the sources of external
potential are fixed nuclei (or pseudopotentials).
For simplicity, we refer to quantum particles as electrons and to
external potentials as nuclei in the rest of this paper.
Our Jastrow factor is applicable to other types of quantum particles
and external potentials.

Any trial wave function can be written in the form
\begin{equation}
\label{eq:generic_wfn}
\Psi_{\rm T}({\bf R}) = e^{J({\bf R})} \Psi_{\rm S}({\bf R}) \;,
\end{equation}
where $\Psi_{\rm S}({\bf R})$ is the part of the wave function that
imposes the symmetry and boundary conditions, and $e^{J({\bf R})}$ is
a Jastrow correlation factor which is constrained so that the symmetry
and boundary properties of $\Psi_{\rm S}({\bf R})$ are transferred
unmodified to $\Psi_{\rm T}({\bf R})$.
In the single-determinant (SD) Slater-Jastrow wave function,
$\Psi_{\rm S}({\bf R})$ is a Slater determinant.
There are various alternatives to Slater determinants, with different
advantages and disadvantages depending on the system.

Typically $J({\bf R})$ is constructed as a sum of terms, e.g.,
\begin{equation}
\label{eq:jastrow_exp_in_terms}
J({\bf R}) = J_{\rm e-e}({\bf R}) + J_{\rm e-n}({\bf R}) +
             J_{\rm e-e-n}({\bf R}) + \ldots \;,
\end{equation}
where ``e--e'' stands for ``electron-electron,'' ``e--n'' for
``electron--nucleus,'' etc.  Each of these terms involves different
numbers of electrons $n$ and nuclei $m$.
We shall refer to $n$ and $m$ as the electronic and nuclear ranks of a
term, respectively, which are constrained to satisfy $n+m\geq 2$,
$n\geq 1$, and $m\geq 0$.
We have designed a generic Jastrow term of selectable ranks,
$J_{n,m}({\bf R})$, such that the total Jastrow factor is constructed
as the exponential of a sum of one or more terms of the desired ranks.
In this notation, $J_{\rm e-e} \equiv J_{2,0}$,
$J_{\rm e-n} \equiv J_{1,1}$, etc.

The function $J_{n,m}({\bf R})$ is a sum over all sets of $n$
electrons and $m$ nuclei in the system of a parameterized function of
the e--e and e--n relative position vectors within each such set.
While alternatives exist, a natural way of parameterizing this
function for arbitrary values of $n$ and $m$ (implying an arbitrary
number of variables in the function) is to expand it in products of
functions of the individual e--e and e--n vectors.
Thus, we construct our Jastrow factor using pairwise objects as
building blocks, and in what follows we describe these objects and
derive the properties of $J_{n,m}({\bf R})$ that follow from those of
the pairwise objects.

We name the e--e functions used in the expansion
$\Phi_\nu^P({\bf r})$, where ${\bf r}$ is the relevant e--e relative
position vector, $\nu$ is the index of the function within a chosen
basis of functions, and $P$ is the e--e dependency index, which allows
the use of different optimizable parameters, if present, for
parallel- and antiparallel-spin electron pairs, for example.
Similarly, the e--n basis functions are $\Theta_\mu^S({\bf r})$, where
${\bf r}$ is the relevant e--n relative position vector, $\mu$ is the
index of the function within the chosen basis set, and $S$ is the e--n
dependency index of the basis set, which allows the use of different
parameters for up- and down-spin electrons around a given nucleus, or
for different atoms, for example.
In the case of nonelectronic systems, e--e and e--n dependency indices
are used to distinguish between particle types and spins.

We introduce a compact notation for defining $J_{n,m}({\bf R})$.
We represent the $n$ electronic indices by the integer vector
$\ivec = \{i_1,i_2,\ldots,i_n\}$, each of whose components takes a
distinct value between $1$ and $N$, and the $m$ nuclear indices by the
integer vector $\Ivec = \{I_1,I_2,\ldots,I_m\}$, each of whose
components takes a distinct value between $1$ and $M$.
For each term in the Jastrow factor we define the e--e and e--n
dependency matrices $\Pmat$ and $\Smat$ of respective sizes
$N\times N$ and $N \times M$ containing the dependency indices
$P_{ij}$ and $S_{iI}$ for each e--e and e--n pair.
The components of $\Pmat$ and $\Smat$ can be made equal depending on
the symmetries of the system, including particle distinguishability
and geometrical symmetries which make different nuclei equivalent.

Likewise, it is convenient to use matrices to represent the basis
functions involved in the Jastrow factor term.
For e--e basis functions, each row and column of the $n\times n$
matrix $\Phimat$ corresponds to an electron,
\begin{widetext}
\begin{equation}
\label{eq:def_phimat}
\Phimat_\numat^\Pmat(\ivec) =
\left(
  \begin{matrix}
    0                                                     &
    \Phi_{\nu_{i_1 i_2}}^{P_{i_1 i_2}}({\bf r}_{i_1 i_2}) &
    \cdots                                                &
    \Phi_{\nu_{i_1 i_n}}^{P_{i_1 i_n}}({\bf r}_{i_1 i_n}) \\
    \Phi_{\nu_{i_1 i_2}}^{P_{i_1 i_2}}({\bf r}_{i_2 i_1}) &
    0                                                     &
    \cdots                                                &
    \Phi_{\nu_{i_2 i_n}}^{P_{i_2 i_n}}({\bf r}_{i_2 i_n}) \\
    \vdots                                                &
    \vdots                                                &
    \ddots                                                &
    \vdots                                                \\
    \Phi_{\nu_{i_1 i_n}}^{P_{i_1 i_n}}({\bf r}_{i_n i_1}) &
    \Phi_{\nu_{i_2 i_n}}^{P_{i_2 i_n}}({\bf r}_{i_n i_2}) &
    \cdots                                                &
    0                                                     \\
  \end{matrix}
\right) \;.
\end{equation}
\end{widetext}
We refer to the $n \times n$ matrix formed by the e--e dependency
indices $\{P_{i_\alpha i_\beta}\}_{\alpha,\beta=1,\ldots,n}$ as
$\Pmat(\ivec)$.
Both $\Pmat(\ivec)$ and the $n\times n$ matrix of e--e expansion
indices $\numat$ are defined to be symmetric, and this fact has been
used in Eq.\ (\ref{eq:def_phimat}).
Noting that ${\bf r}_{ji}=-{\bf r}_{ij}$, and restricting the e--e
functions to be either symmetric or antisymmetric about the origin,
one finds in Eq.\ (\ref{eq:def_phimat}) that matrix $\Phimat$ is
symmetric, antisymmetric, or asymmetric depending on whether the
functions in the basis set are all symmetric, all antisymmetric, or
both types are present, respectively.

For e--n basis functions each row of the $n\times m$ matrix
$\Thetamat$ corresponds to an electron and each column to a nucleus,
\begin{equation}
\label{eq:def_thetamat}
\Thetamat_\mumat^\Smat(\ivec,\Ivec) =
\left(
  \begin{matrix}
    \Theta_{\mu_{i_1 I_1}}^{S_{i_1 I_1}}({\bf r}_{i_1 I_1}) &
    \cdots                                                  &
    \Theta_{\mu_{i_1 I_m}}^{S_{i_1 I_m}}({\bf r}_{i_1 I_m}) \\
    \vdots                                                  &
                                                            &
    \vdots                                                  \\
    \Theta_{\mu_{i_n I_1}}^{S_{i_n I_1}}({\bf r}_{i_n I_1}) &
    \cdots                                                  &
    \Theta_{\mu_{i_n I_m}}^{S_{i_n I_m}}({\bf r}_{i_n I_m}) \\
  \end{matrix}
\right) \;.
\end{equation}
We refer to the $n \times m$ matrix formed by the e--n dependency
indices $\{S_{i_\alpha I_\beta}\}_{\alpha=1,\ldots,n;
~\beta=1,\ldots,m}$ as $\Smat(\ivec,\Ivec)$, and the $n \times m$
matrix of e--n expansion indices is $\mumat$.

We write $J_{n,m}$ as a sum of contributions from each group of $n$
electrons and $m$ nuclei in the system,
\begin{equation}
\label{eq:gjastrow_sum_of_core}
J_{n,m} =
  \frac{1}{n! m!} \sum_\ivec \sum_\Ivec J_{n,m}(\ivec,\Ivec) =
  \sum_\ivec^{\rm s.v.} \sum_\Ivec^{\rm s.v.} J_{n,m}(\ivec,\Ivec) \;,
\end{equation}
where summations with vector indices represent sums in which every
component of the vector is a summation index, and ``$\rm s.v.$'' (for
``sorted vector'') indicates that the sum is restricted to vectors
whose components are sorted, e.g., $i_1<i_2<\ldots<i_n$, which avoids
redundant contributions \cite{footnote1}.
The contribution from the $n$-electron and $m$-nucleus group
$\{\ivec,\Ivec\}$ is
\begin{equation}
\label{eq:gjastrow_core}
J_{n,m}(\ivec,\Ivec) =
  \sum_\numat^{\rm u.t.} \sum_\mumat
  \lambda_{\numat, \mumat}^{\Pmat(\ivec), \Smat(\ivec,\Ivec)}
  \prod^{\rm u.t.} \Phimat_\numat^\Pmat(\ivec)
  \prod \Thetamat_\mumat^\Smat(\ivec,\Ivec) \;,
\end{equation}
where $\lambda$ are the linear parameters, summations with matrix
indices represent sums in which every component of the matrix is a
summation index, $\prod{}$ acting on matrices implies the product of
all of their components, and ``u.t.'' means that the relevant
operation is restricted to the upper-triangular portion of the e--e
matrices involved, excluding the diagonal.

\subsection{Symmetry properties of the linear parameters}

Equation (\ref{eq:gjastrow_sum_of_core}) imposes the condition that
$J_{n,m}(\ivec,\Ivec)$ must not depend on the specific ordering of
the electrons and nuclei listed in $\ivec$ and $\Ivec$.
Let $\Amat$ and $\Bmat$ be permutation matrices of respective sizes
$n\times n$ and $m\times m$ such that $\Amat\MV\ivec$ and
$\Bmat\MV\Ivec$ are integer vectors containing reordered electronic
and nuclear indices.
The value of $J_{n,m}(\Amat\MV\ivec,\Bmat\MV\Ivec)$ should therefore
equal that of $J_{n,m}(\ivec,\Ivec)$,
\begin{widetext}
\begin{eqnarray}
\label{eq:gjastrow_core_permute}
\nonumber
J_{n,m}(\Amat\MV\ivec,\Bmat\MV\Ivec) & = &
 \sum_{\numat(\Amat\MV\ivec)}^{\rm u.t.}
 \sum_{\mumat(\Bmat\MV\Ivec)}
 \lambda_{\numat(\Amat\MV\ivec), \mumat(\Amat\MV\ivec,\Bmat\MV\Ivec)}^{
          \Pmat (\Amat\MV\ivec), \Smat (\Amat\MV\ivec,\Bmat\MV\Ivec)}
 \prod^{\rm u.t.} \Phimat_\numat^\Pmat  (\Amat\MV\ivec)
 \prod            \Thetamat_\mumat^\Smat(\Amat\MV\ivec,\Bmat\MV\Ivec)
\\ & = & \nonumber
 \sum_{\Amat\MM\numat\MM\AmatT}^{\rm u.t.}
 \sum_{\Amat\MM\mumat\MM\BmatT}
 \lambda_{\Amat\MM\numat\MM\AmatT, \Amat\MM\mumat\MM\BmatT}^{
          \Amat\MM\Pmat(\ivec) \MM\AmatT, \Amat\MM\Smat(\ivec,\Ivec) \MM\BmatT}
 \prod^{\rm u.t.} \Amat\MM\Phimat_\numat^\Pmat(\ivec)\MM\AmatT
 \prod            \Amat\MM\Thetamat_\mumat^\Smat(\ivec,\Ivec)\MM\BmatT
\\ & = &
 \sum_{\numat}^{\rm u.t.}
 \sum_{\mumat}
 \lambda_{\Amat\MM\numat\MM\AmatT, \Amat\MM\mumat\MM\BmatT}^{
          \Amat\MM\Pmat(\ivec) \MM\AmatT, \Amat\MM\Smat(\ivec,\Ivec) \MM\BmatT}
 \gamma\left[ \Phimat_\numat^\Pmat(\ivec),\Amat \right]
 \prod^{\rm u.t.} \Phimat_\numat^\Pmat(\ivec)
 \prod            \Thetamat_\mumat^\Smat(\ivec,\Ivec) \;,
\end{eqnarray}
\end{widetext}
where
\begin{equation}
\gamma\left[ \Phimat_\numat^\Pmat(\ivec),\Amat \right] =
  \frac{\prod^{\rm u.t.}\Amat\MM\Phimat_\numat^\Pmat(\ivec)\MM\AmatT}
       {\prod^{\rm u.t.}\Phimat_\numat^\Pmat(\ivec)} \;,
\end{equation}
which is $+1$ for basis sets consisting only of symmetric functions,
while in the presence of antisymmetric basis functions it may be $+1$
or $-1$ depending on the precise permutation performed by $\Amat$.
Equating the right-hand sides of Eqs.\ (\ref{eq:gjastrow_core}) and
(\ref{eq:gjastrow_core_permute}) one finds that
\begin{equation}
\label{eq:lambda_symmetry}
\lambda_{\numat,\mumat}^{\Pmat(\ivec),\Smat(\ivec,\Ivec)} =
\gamma\left[ \Phimat_\numat^\Pmat(\ivec),\Amat \right]
\lambda_{\Amat\MM\numat\MM\AmatT,
         \Amat\MM\mumat\MM\BmatT}^{
         \Amat\MM\Pmat(\ivec)\MM\AmatT,
         \Amat\MM\Smat(\ivec,\Ivec)\MM\BmatT} \;.
\end{equation}
This equation represents the basic symmetry property of the linear
parameters of the Jastrow factor, which implies that a parameter with
a given set of superindices $\{\Pmat(\ivec),\Smat(\ivec,\Ivec)\}$ is
determined by another parameter with a permuted set of superindices
$\{\Amat\MM\Pmat(\ivec)\MM\AmatT,
\Amat\MM\Smat(\ivec,\Ivec)\MM\BmatT\}$.
This redundancy is removed by considering only one of the possible
permutations of $\{\Pmat(\ivec),\Smat(\ivec,\Ivec)\}$.
We call this particular permutation of $\{\Pmat(\ivec),
\Smat(\ivec,\Ivec)\}$ the signature of the group of particles
$\{\ivec,\Ivec\}$,
\begin{equation}
\{\Ptilde(\ivec),\Stilde(\ivec,\Ivec)\}=
\{\Umat\MM\Pmat(\ivec)\MM\UmatT,\Umat\MM\Smat(\ivec,\Ivec)\MM\RmatT\} \;,
\end{equation}
where the permutation matrices $\{\Umat,\Rmat\}$ are computed by
applying a matrix-sorting algorithm \cite{footnote2} to
$\{\Pmat(\ivec),\Smat(\ivec,\Ivec)\}$.
In our terminology, the set of linear parameters whose superindices
reduce to the same signature constitute a parameter channel.
Only those parameters whose superindices equal the signature of a
channel need be stored, and any other linear parameters in the channel
can be computed from them via Eq.\ (\ref{eq:lambda_symmetry}).

The signature $\{\Ptilde(\ivec),\Stilde(\ivec,\Ivec)\}$ may contain
repeated entries such that there exist permutation matrices
$\{\Amat,\Bmat\}$ that leave the signature unchanged,
\begin{equation}
\{\Ptilde(\ivec),\Stilde(\ivec,\Ivec)\}=
\{\Amat\MM\Ptilde(\ivec)\MM\AmatT,\Amat\MM\Stilde(\ivec,\Ivec)\MM\BmatT\} \;,
\end{equation}
in which case Eq.\ (\ref{eq:lambda_symmetry}) becomes
\begin{equation}
\label{eq:lambda_symmetry_invariant_signature}
\lambda_{\numat,\mumat}^{\Ptilde(\ivec),\Stilde(\ivec,\Ivec)} =
\gamma\left[ \Phimat_\numat^\Ptilde(\ivec),\Amat \right]
\lambda_{\Amat\MM\numat\MM\AmatT,
         \Amat\MM\mumat\MM\BmatT}^{\Ptilde(\ivec),\Stilde(\ivec,\Ivec)} \;.
\end{equation}
Equation (\ref{eq:lambda_symmetry_invariant_signature}) is the
symmetry constraint that relates linear parameters within a channel,
which can be imposed as detailed in Sec.~\ref{sec:constraints}.

\subsection{Indexing of basis functions}

The components of $\numat$ and $\mumat$ are the e--e and e--n
expansion indices.  We define the expansion indices so that they can
each take any value between 1 and the e--e expansion order $p$, and
between 1 and the e--n expansion order $q$, respectively.
We factorize an optional cutoff function into $\Phi_\nu^P$ and
$\Theta_\mu^S$, so that
\begin{equation}
\label{eq:Phi_factored}
\Phi_\nu^P({\bf r}) = f^P({\bf r}) \phi_\nu^P({\bf r}) \;,
\end{equation}
for $\nu>0$, and
\begin{equation}
\Theta_\mu^S({\bf r}) = g^S({\bf r}) \theta_\mu^S({\bf r}) \;,
\end{equation}
for $\mu>0$, where $f^P$ and $g^S$ are the e--e and e--n cutoff
functions and $\phi_\nu^P$ and $\theta_\mu^S$ are functions from a
suitable basis set.
This factorization allows an efficient implementation of localized
Jastrow factor terms.

Additionally, we allow expansion indices to take a value of zero with
the special meaning that $\Phi_0^P({\bf r})=\Theta_0^S({\bf r})=1$
for all $P$, $S$, and ${\bf r}$.  Note that these 0-th functions do
not contain cutoff functions.  This allows us to construct terms with
specialized functional forms, such as those involving dot products of
vectorial quantities.

\subsection{Constraints}
\label{sec:constraints}

Constraints on the parameters can be expressed in the form of a system
of equations involving the linear parameters and the basis function
parameters.
We restrict our analysis to linear constraints on the linear
parameters, and constraints that can be imposed on the nonlinear
parameters contained in a basis function independently from the linear
parameters and from nonlinear parameters in other basis functions.

Linear constraints on the linear parameters can be imposed using
Gaussian elimination, as described in
Ref.~\cite{drummond_jastrow_2004}.
The matrix of coefficients may depend on the nonlinear parameters in
the basis functions, if present, and the linear system is usually
underdetermined, resulting in a subset of the parameters being
determined by the values of the remaining parameters, which can be
optimized directly.

When a constraint results in setting specific linear parameters to
zero, it is more convenient simply to remove them from the list of
linear parameters.
This is accomplished by disallowing the indices $\numat$ and $\mumat$
from taking the values corresponding to linear parameters.
We call this an indexing constraint.

\subsubsection{Symmetry and antisymmetry constraints}

Symmetry constraints must always be imposed, otherwise the trial wave
function is unphysical and calculations give erroneous results.
Symmetry constraints amount to equalities between pairs of parameters
as per Eq.\ (\ref{eq:lambda_symmetry_invariant_signature}).
When two of these equalities relate the same pair of parameters
with opposite signs, e.g., $\lambda_1=\lambda_2$ and
$\lambda_1=-\lambda_2$, which implies $\lambda_1=\lambda_2=0$, both
parameters are eliminated using indexing constraints.

\subsubsection{Constraints at e--e and e--n coalescence points}

The Coulomb potential energy diverges when the positions of two
electrons or an electron and a nucleus coincide.
However, the local energy of an eigenstate of the Hamiltonian,
including the exact ground-state wave function, is finite and
constant throughout configuration space.
Divergences in the local energy are therefore not a feature of the
exact wave function and can lead to poor statistics in QMC
calculations; hence it is important to avoid them.
The kinetic energy must diverge to cancel out the potential energy
and keep the local energy finite, which is achieved by demanding that
the wave function obeys the Kato cusp conditions
\cite{kato_eigenfunctions_1957}.
For any two charged particles $i$ and $j$ in a two- or
three-dimensional system, these are
\begin{equation}
\label{eq:kato}
\left(
 \frac{1}{\Psi}
 \frac{\partial {\hat\Psi}}
      {\partial r_{ij}}
\right)_{r_{ij}\rightarrow 0} =
 \frac {2 q_i q_j \mu_{ij} } {d \pm 1} \;,
\end{equation}
where ${\hat\Psi}$ denotes the spherical average of $\Psi$,
$q$ represents charge, $\mu_{ij}=m_i m_j / (m_i + m_j)$ is
the reduced mass, $m$ represents mass, $d$ is the dimensionality, and
the positive sign in the denominator is for indistinguishable
particles and the negative sign is for distinguishable particles.
Fixed nuclei are regarded as having an infinite mass.
Divergent interactions other than the Coulomb potential would give
rise to different expressions.

It is common practice to impose the e--n cusp conditions on
$\Psi_{\rm S}$ and the e--e cusp conditions on the Jastrow factor.
This is because typical forms of $\Psi_{\rm S}$ explicitly depend on
the e--n distances but not on the e--e distances.
Our implementation allows the option of applying both types of cusp
conditions to the Jastrow factor, which gives flexibility in the
choice of $\Psi_{\rm S}$ and its properties.
In particular, we impose the cusp conditions on a single Jastrow
factor term, and constrain all other terms in the Jastrow factor so
that their contribution to the local kinetic energy is finite at e--e
and e--n coalescence points.
For nondivergent interaction potentials, such as most
pseudopotentials, we simply require that the kinetic energy remains
finite at coalescence points.
Our implementation is also capable of not applying any constraints at
e--e and e--n coalescence points since this is advantageous in some
cases \cite{per_anisotropic_2009,per_personal_2012}.

Imposing that the kinetic energy be finite at coalescence points is
nontrivial if the Jastrow factor contains anisotropic functions.
Consider the exponent of a Jastrow factor $J$ near a point where two
particles coalesce, be it two electrons or an electron and a nucleus.
The dependence of $J$ on coordinates other than those of the
coalescing particles should be smooth in the vicinity of the
coalescence point, and therefore one should be able to write
$J=J({\bf r})$, where ${\bf r}$ is the difference between the
position vector of the two particles, and all remaining particles are
held fixed.

The local kinetic energy is computed from two estimators, one
involving ${\bm\nabla} J({\bf r})$ and the other
$\nabla^2 J({\bf r})$.
We require both quantities to remain finite as
${\bf r}\rightarrow {\bm 0}$.
We expand the Jastrow factor in spherical harmonics,
$J({\bf r}) = \sum_{l=0}^{\infty} \sum_{m=-l}^l J^{(l,m)}$, with
$J^{(l,m)} = f_{l,m}(r)Y_{l,m}(\theta,\phi)$, and the gradient and
Laplacian of $J^{(l,m)}$ are
\begin{eqnarray}
\label{eq:def_grad_J}
{\bm\nabla} J^{(l,m)} & = &
  f_{l,m}^\prime(r) Y_{l,m}(\theta,\phi) {\bf u}_r
\nonumber \\ & & +
  \frac{f_{l,m}(r)}{r}
  \frac{\partial Y_{l,m}(\theta,\phi)}{\partial \theta}
  {\bf u}_\theta
\nonumber \\ & & +
  \frac{f_{l,m}(r)}{r\sin\theta}
  \frac{\partial Y_{l,m}(\theta,\phi)}{\partial \phi}
  {\bf u}_\phi \;,
\end{eqnarray}
\begin{eqnarray}
\label{eq:def_lap_J}
\nabla^2 J^{(l,m)} & = &
  \left[
    f_{l,m}^{\prime\prime}(r)
  +
    \frac{2f_{l,m}^{\prime}(r)}{r}
\right. \nonumber \\ & & \left. -
    \frac{l(l+1)f_{l,m}(r)}{r^2}
  \right] Y_{l,m}(\theta,\phi) \;.
\end{eqnarray}
Let us assume that $f_{l,m}(r)$ is finite at the origin and expand it to
second order about $r=0$, $f_{l,m}(r) \approx f_{l,m}(0) +
f_{l,m}^\prime(0)r + f_{l,m}^{\prime\prime}(0)r^2/2$.  Substituting into
Eqs.\ (\ref{eq:def_grad_J}) and (\ref{eq:def_lap_J}), and ignoring
contributions of $\mathcal{O}(r)$ or higher, we arrive at
\begin{eqnarray}
\label{eq:grad_J}
{\bm\nabla} J^{(l,m)} & \approx &
  f_{l,m}^\prime(0) Y_{l,m}(\theta,\phi) {\bf u}_r
\nonumber \\ & & +
  \left[
    \frac{f_{l,m}(0)}{r} + f_{l,m}^\prime(0)
  \right] \frac{\partial Y_{l,m}(\theta,\phi)}{\partial \theta} {\bf u}_\theta
\nonumber \\ & & +
  \left[
    \frac{f_{l,m}(0)}{r\sin\theta} +
    \frac{f_{l,m}^\prime(0)}{\sin\theta}
  \right] \frac{\partial Y_{l,m}(\theta,\phi)}{\partial \phi} {\bf u}_\phi \;,
\end{eqnarray}
\begin{eqnarray}
\label{eq:lap_J}
\nabla^2 J^{(l,m)} & \approx &
  \left[
    \left(3-\frac{l(l+1)}{2} \right) f_{l,m}^{\prime\prime}(0)
\nonumber \right. \\ & & + \left.
    \left(2-l(l+1)\right)\frac{f_{l,m}^{\prime}(0)}{r}
\nonumber \right. \\ & & - \left.
    l(l+1)\frac{f_{l,m}(0)}{r^2}
  \right] Y_{l,m}(\theta,\phi) \;.
\end{eqnarray}
The coefficient of the negative powers of $r$ in
Eqs.\ (\ref{eq:grad_J}) and (\ref{eq:lap_J}) must vanish for
${\bm\nabla}J$ and $\nabla^2 J$ to be finite at the coalescence
point.
This gives rise to two conditions:
(a) if $l\neq 0$ then $f_{l,m}(0)=0$; and
(b) if $l\neq 1$ then $f_{l,m}^\prime(0)=0$.

Application of the Kato cusp or finite kinetic energy constraints
requires the construction of a linear system for each linear-parameter
channel in each term of the Jastrow factor based on the above
equations.
Let ${\mathcal P}_{l,m}$ be an operator such that
${\mathcal P}_{l,m} \sum_{l=0}^{\infty} \sum_{m=-l}^l
f_{l,m}(r) Y_{l,m}(\theta,\phi) = f_{l,m}(r)$.
The cusp equations associated with the coalescence of electrons $i$
and $j$ have the form
\begin{equation}
\label{eq:ee_cusp}
\sum_{\numat}^{{\rm u.t.}, {\rm o.c.}}
\sum_{\mumat}^{{\rm n.c.}}
  \lambda_{\numat, \mumat}^{\Pmat, \Smat}
  \left[
    \frac {\partial {\mathcal P}_{0,0}\left[
           \Phi_{\nu_{ij}}^{P_{ij}}({\bf r}_{ij})\right]}
          {\partial r_{ij}}
  \right]_{r_{ij}=0} = \Gamma_{ij} \;,
\end{equation}
where $\Gamma_{ij}$ is the right-hand side of Eq.\ (\ref{eq:kato}).
The label ``${\rm o.c.}$'' (for ``one contribution'') denotes that the
sum is restricted to values of the $\numat$ sets such that the
elements of the upper triangular portion of
$\Phimat_\numat^\Pmat(\ivec)$ are all 1 except that corresponding to
the electron pair formed by $i$ and $j$, and the label
``${\rm n.c.}$'' (for ``no contribution'') denotes that the sum is
restricted to values of $\mumat$ such that all elements of
$\Thetamat_\mumat^\Smat(\ivec,\Ivec)$ are 1.
These restrictions are trivially satisfied by e--e terms.

Parameters that do not contribute to Eq.\ (\ref{eq:ee_cusp}) should be
set by the condition that the kinetic energy does not diverge at
e--e coalescence points, resulting in
\begin{equation}
\label{eq:ee_noaniso}
\sum_{\numat}^{{\rm u.t.}, {\rm e.p.}}
\sum_{\mumat}^{{\rm e.p.}}
  \lambda_{\numat, \mumat}^{\Pmat, \Smat}
  \left[
    {\mathcal P}_{l,m}\left[\Phi_{\nu_{ij}}^{P_{ij}}({\bf r}_{ij})\right]
    \right]_{r_{ij}=0} = 0 \;,
\end{equation}
for $l\neq 0$, and
\begin{equation}
\label{eq:ee_nocusp}
\sum_{\numat}^{{\rm u.t.}, {\rm e.p.}}
\sum_{\mumat}^{{\rm e.p.}}
  \lambda_{\numat, \mumat}^{\Pmat, \Smat}
  \left[
    \frac {\partial {\mathcal P}_{l,m}\left[
           \Phi_{\nu_{ij}}^{P_{ij}}({\bf r}_{ij})\right]}
          {\partial r_{ij}}
  \right]_{r_{ij}=0} = 0 \;,
\end{equation}
for $l\neq 1$.
The anisotropy of $\Phi_{\nu_{ij}}^{P_{ij}}({\bf r}_{ij})$ at
${\bf r}_{ij}={\bm 0}$ determines which of Eqs.\ (\ref{eq:ee_noaniso})
and Eqs.\ (\ref{eq:ee_nocusp}) need be imposed.
The label ${\rm e.p.}$ (for ``equal-product'') denotes that the sum
is only over indices associated with electrons $i$ and/or $j$, and
these indices take only values such that the product of the pair of
functions associated with $\nu_{ik}$ and $\nu_{jk}$ ($\mu_{iI}$ and
$\mu_{jI}$ in the e--n case) is equal throughout the sum.
Each set of distinct two-function products and each value of
$(l,m)$ correspond to different equations, and each set of possible
values of the indices not summed over corresponds to a separate set
of equations.

For the coalescence of electron $i$ and nucleus $I$ the cusp
conditions take the form
\begin{equation}
\label{eq:en_cusp}
\sum_{\numat}^{{\rm u.t.}, {\rm n.c.}}
\sum_{\mumat}^{{\rm o.c.}}
  \lambda_{\numat, \mumat}^{\Pmat, \Smat}
  \left[
    \frac {\partial {\mathcal P}_{l,m}\left[
           \Theta_{\mu_{iI}}^{S_{iI}}({\bf r}_{iI})\right]}
          {\partial r_{iI}}
  \right]_{r_{iI}=0} = \Gamma_{iI} \;,
\end{equation}
while the kinetic energy is kept finite if
\begin{equation}
\label{eq:en_noaniso}
\sum_{\numat,\mumat}^{{\rm u.t.}, {\rm e.p.}}
  \lambda_{\numat, \mumat}^{\Pmat, \Smat}
  \left[
    {\mathcal P}_{l,m}\left[\Theta_{\mu_{iI}}^{S_{iI}}({\bf r}_{iI})\right]
  \right]_{r_{iI}=0} = 0 \;,
\end{equation}
for $l\neq 0$ and
\begin{equation}
\label{eq:en_nocusp}
\sum_{\numat,\mumat}^{{\rm u.t.}, {\rm e.p.}}
  \lambda_{\numat, \mumat}^{\Pmat, \Smat}
  \left[
    \frac {\partial {\mathcal P}_{l,m}\left[
           \Theta_{\mu_{iI}}^{S_{iI}}({\bf r}_{iI})\right]}
          {\partial r_{iI}}
  \right]_{r_{iI}=0} = 0 \;,
\end{equation}
for $l\neq 1$.
The equal-product constraint on the sum is now such that the
sum is only over e--e indices associated with electron
$i$ and e--n indices associated with nucleus $I$, and
these indices only take values such that the product of the pair of
functions associated with $\nu_{ik}$ and $\mu_{kI}$ is equal throughout
the sum.  Again, each set of distinct two-function products corresponds
to a different equation, and each set of possible values of the indices
not being summed over corresponds to a separate set of equations.

Note that the equal-product constraints in the sums of
Eqs.\ (\ref{eq:ee_cusp}) and (\ref{eq:en_cusp}) reduce to the equal-product
constraints described for the e--e--n $f$ term of
the Drummond-Towler-Needs (DTN) Jastrow factor in the Appendix of
Ref.~\cite{drummond_jastrow_2004} when natural-power basis
functions are chosen.

\subsubsection{Other constraints}

It is possible to construct terms containing dot products by using
appropriate constraints.
For example, consider the basis functions $\Theta_1({\bf r})=x$,
$\Theta_2({\bf r})=y$, and $\Theta_3({\bf r})=z$.
In an e--n--n term we can restrict the indices so that $\mumat$ takes
only the values $\left(1~1\right)$, $\left(2~2\right)$,
$\left(3~3\right)$, so that the contribution of electron $i$
and nuclei $I$ and $J$ is ${\bf r}_{i I}\cdot{\bf r}_{i J}$, provided
we also apply a linear constraint that equates the three nonzero
linear coefficients.
Section \ref{subsec:specific_terms} gives a practical example of
a term containing dot products, which is used in
Sec.~\ref{subsec:h2_mol}.

It is also possible to introduce Boys-Handy-style indexing,
\cite{boys_calculation_1969} where the sum of all e--e and e--n
indices is restricted to be less than or equal to some fixed integer
$l$.
This is accomplished by setting the e--e and e--n expansion orders
to $l$ and then eliminating the parameters that violate the
conditions via indexing restrictions.

\section{Basis functions and terms}
\label{sec:construction}

\subsection{Basis sets and cutoff functions}

Possibly the simplest basis set is the natural powers,
\begin{equation}
N_\nu({\bf r})=r^{\nu-1} \;,
\end{equation}
as used in the DTN Jastrow factor for the localized $u$, $\chi$, and
$f$ terms \cite{drummond_jastrow_2004}.
These functions need to be cut off at some radius $L$, for which
purpose the DTN Jastrow factor uses the polynomial
cutoff function
\begin{equation}
D({\bf r})=(r-L)^C \Theta(L-r) \;,
\end{equation}
where $L$ is an optimizable parameter, $C$ is a positive integer, and
$\Theta(r)$ is the Heaviside step function.
We also use a slightly different version of this cutoff function,
\begin{equation}
P({\bf r})=(1-r/L)^C \Theta(L-r) \;,
\end{equation}
which should be numerically superior to $D({\bf r})$.

A particular variant of $P({\bf r})$ is the anisotropic
cutoff function
\begin{equation}
A({\bf r}) =
  (1-r/L)^C \Theta(L-r)
  \sum_i c_i
    \prod_\beta^d
      \left[
        \frac {{\bf r}\cdot {\hat {\bf u} }_\beta}
              {r}
      \right]^{p^{(i)}_\beta} \;,
\end{equation}
where $L$ is an optimizable parameter, $C$ is a positive integer, $d$
is the dimensionality of the system,
$\{{\hat {\bf u} }_\beta\}_{\beta=1,\ldots,d}$ are unit vectors along
$d$ orthogonal directions, $\{c_i\}$ are real-valued constants, and
$p_\beta^{(i)}$ are integer exponents, which are constrained so that
$\sum_\beta^d p_\beta^{(i)}$ is the same for all values of
$i$.  This cutoff function is simply the product of an isotropic
cutoff function and a spherical harmonic.
For example, with $c_1=3$, $c_2=-1$, ${\bf p}^{(1)}=(2~1~1)$, and
${\bf p}^{(2)}=(0~3~1)$, and the vectors pointing along the
Cartesian axes, we obtain
\begin{equation}
A({\bf r}) =
  (1-r/L)^C \Theta(L-r)
  \left[ \frac {(3x^2-y^2)yz} {r^4} \right] \;,
\end{equation}
which is proportional to a real spherical harmonic with $l=4$.  The
advantage of describing anisotropy in the cutoff function rather than
in the basis functions is that the common spherical harmonic can be
factorized out of the sum over expansion indices, which reduces the
computational cost.
We allow different orientations to be used for different e--e or e--n
dependency indices, which is useful to adapt the functional form to,
e.g., the geometry of a molecule.

An alternative to the natural-power basis in finite systems
is a basis of powers of fractions which tend to a constant
as $r\rightarrow\infty$, and therefore do not need to be cut off.
We define the basis
\begin{equation}
F_\nu({\bf r})=\left(\frac{r}{r^b+a}\right)^{\nu-1} \;,
\end{equation}
where $a$ and $b$ are real-valued optimizable parameters.
Similar basis sets with $b=1$ have been used in the
literature, often in conjunction with Boys-Handy-style
indexing
\cite{boys_calculation_1969,schmidt_correlated_1990,filippi_multiconfiguration_1996,per_anisotropic_2009}, and this
basis was used in Ref.~\cite{seth_quantum_2011} with an early
implementation of the Jastrow factor presented here.

In extended systems it is important to use a basis that is consistent
with the geometry of the simulation cell and has the periodicity of
the system, such as a cosine basis,
\begin{equation}
C_\nu({\bf r}) = \sum_{{\bf G} \in
\nu\textrm{-th~star}} \cos\left( {\bf G} \cdot {\bf r} \right) \;,
\end{equation}
where the $\bf G$ vectors are arranged in stars defined by the cell
geometry.
This basis is used in the DTN Jastrow factor for the extended $p$ and
$q$ terms.

A suitable basis set for building specialized terms containing
dot products is
\begin{equation}
V_\nu({\bf r}) =
  r^{{\rm INT}\left[\left(\nu-1\right)/d\right]}
    \frac {{\bf r} \cdot {\hat{\bf u}_{{\rm MOD}\left(\nu-1,d\right)+1}}}
          {r} \;,
\end{equation}
where $d$ is the dimensionality of the system and
$\{{\hat{\bf u}}_\beta\}_{\beta=1,\ldots,d}$ are unit vectors parallel
to the $d$ Cartesian axes.
A term constructed using these functions with appropriate
index-restriction constraints would consist of dot products between
two vectors multiplied by a natural-power expansion in their moduli.

\subsection{Terms and notation}
\label{subsec:specific_terms}

We employ a condensed notation to refer to Jastrow terms
that use certain basis functions, cutoff functions and constraints.
Each term is represented by a single capital letter, with $n$ and $m$
as subindices.
Any other relevant information is given as a superindex.
Typically we use expansion orders $p$ and $q$ of 7--9 for two-body
terms, 4--5 for three-body terms, and 2--3 for four-body terms,
except when indicated otherwise.

For simple Jastrow terms we use the natural power basis functions
$N_\nu$ and the polynomial cutoff functions $P$ or $D$.
We refer to these terms as $N_{n,m}$.
$N_{2,0}$, $N_{1,1}$, and $N_{2,1}$ are the equivalent of the DTN $u$,
$\chi$, and $f$ terms, respectively.
In the $N_{2,1}$ term, and in any term where more than one electron
and one or more nuclei are involved, we choose not to apply e--e
cutoff functions, relying instead on the e--n cutoffs to fulfill
this role.
Additional $N_{n,m}$ terms used in this paper that were not part
of the DTN Jastrow factor are $N_{1,2}$, $N_{3,0}$, $N_{1,3}$,
$N_{2,2}$, $N_{3,1}$, and $N_{4,0}$.  In $N_{n,m}$ we typically use a
truncation order in the cutoff function of $C=3$.

We use $A_{n,m}^{\rm s.h.}$ to refer to the anisotropic variant of
$N_{n,m}$.  The $A_{n,m}^{\rm s.h.}$ term consists of natural power
basis functions $N_\nu$ and the anisotropic cutoff function $A$, and
``${\rm s.h.}$'' is a placeholder for describing the spherical
harmonic.  For example, for the highly anisotropic N${}_2$ molecule
we use terms such as $A_{1,1}^{z}$, $A_{1,1}^{z^2}$, $A_{2,1}^{z}$,
and $A_{2,1}^{z^2}$.

In finite systems we also use the $F_\nu$ basis functions in terms
without explicit cutoff functions which we call $F_{n,m}$, or
$F_{n,m}^{b=1}$ when we force $b=1$ in the basis functions.
In some systems it is useful to apply Boys-Handy-style indexing
to $F_{n,m}^{b=1}$, and we refer to the resulting term as
$B_{n,m}$.

In extended systems we make use of the cosine basis functions $C_\nu$
in terms denoted $C_{n,m}$, where we choose expansion orders so that
at least as many ${\bf G}$ vectors as electrons in each
spin channel are included in the expansion.

To test the flexibility of our implementation we have designed an
e--e--n--n Jastrow term for describing the correlations associated
with van der Waals interactions, which we call $V_{2,2}$.
This term is capable of distinguishing between configurations
where the electron-nucleus relative position vectors
${\bf r}_{iI}$ and ${\bf r}_{jJ}$ are parallel from those where
they are antiparallel.
Introducing a dot product achieves this effect, and $V_{2,2}$ has
the following functional form,
\begin{eqnarray}
\nonumber
V_{2,2} & = &
  \frac 1 2
  \sum_{i \neq j}^N \sum_{I \neq J}^M
    P(r_{iI}) P(r_{jJ})
    \sum_{\nu_{ij}}^p
    \sum_{\mu_{iI},\mu_{jJ}}^q
      \lambda_{\nu_{ij}\mu_{iI}\mu_{iJ}}
\\ & & \times
      N_{\nu_{ij}}(r_{ij})
      N_{\mu_{iI}}(r_{iI}) N_{\mu_{jJ}}(r_{jJ})
      {\bf r}_{iI}\cdot{\bf r}_{jJ} \;.
\end{eqnarray}
We require basis functions to be scalars in our Jastrow factor, so the
dot product is separated into its components.
Hence, we construct the $V_{2,2}$ term using $V_\nu$ for the e--n
basis with $P$ as the e--n cutoff functions, and $N_\nu$ for the e--e
basis.
We allow e--n indices to be zero, and apply a number of constraints
on the linear parameters:
(a) we eliminate all index sets except those in which the
e--n indices are of the form
$\mumat =
  \left(
    \begin{matrix}
      k & 0 \\
      0 & l
    \end{matrix}
  \right)
$
or
$ \left(
    \begin{matrix}
      0 & k \\
      l & 0
    \end{matrix}
  \right)
$, with $k,l>0$;
(b) we eliminate all parameters that do not satisfy
${\rm MOD}(k-1,3)={\rm MOD}(l-1,3)$;
(c) we equate any two linear parameters
$\lambda_{\nu,\mumat_1}$ and $\lambda_{\nu,\mumat_2}$ if
the zeros of $\mumat_1$ and $\mumat_2$ are in the same position
and their nonzero components satisfy
${\rm INT}\left[(k_1-1)/3)\right] = {\rm INT}\left[(k_2-1)/3\right]$
and
${\rm INT}\left[(l_1-1)/3)\right] = {\rm INT}\left[(l_2-1)/3\right]$.
These constraints are applied in addition to the generic symmetry and
cusp constraints.
Table~\ref{tab:notation} summarizes the notation for the terms
we have introduced.

\begin{table*}
  \caption{Notation for Jastrow terms correlating $n$ electrons and $m$
           nuclei using different basis functions.
           \label{tab:notation}}
\begin{tabular}{llll}
\hline
\hline
Name                 & Basis set
                     & Cutoff function
                     & Special constraints \\
\hline
$N_{n,m}$            & Natural powers
                     & Polynomial
                     & None \\
$F_{n,m}$            & Powers of $r/(r^b+a)$
                     & None
                     & None \\
$B_{n,m}$            & Powers of $r/(r+a)$
                     & None
                     & Boys-Handy-style indexing \\
$A_{n,m}^{\rm s.h.}$ & Natural powers
                     & Anisotropic polynomial
                     & None \\
$C_{n,m}$            & Cosines
                     & None
                     & None \\
$V_{n,m}$            & Natural powers times unit vectors
                     & Polynomial
                     & Dot product \\
\hline
\hline
\end{tabular}
\end{table*}

\section{Results}
\label{sec:results}

In the present work we have used a variety of methods to optimize
our Jastrow factors, namely variance minimization, minimization of the
mean absolute deviation of the local energy with respect to the median
energy, and linear least-squares energy
minimization \cite{toulouse_emin_2007,umrigar_emin_2007}.
All of our final wave functions are energy-minimized except where
otherwise stated.
Starting with the Hartree-Fock (HF) wave function, we progressively
introduce Jastrow terms and re-optimize all of the parameters
simultaneously.
Optimizing the Jastrow factor term-by-term is unnecessary in practical
applications, but here it allows us to understand the importance of
the different terms.
We refer to the total number of optimizable parameters in the wave
function as $N_p$.

The correlation energy is defined as the difference between the HF
energy and the exact energy, $E_{\rm HF}-E_{\rm exact}$.
The fraction of the correlation energy retrieved in a VMC calculation
with a given trial wave function $\Psi$,
\begin{equation}
f_{\rm CE}[\Psi] =
  \frac{E_{\rm HF}-E_{\rm VMC}[\Psi]}
       {E_{\rm HF}-E_{\rm exact}} \;,
\end{equation}
is a measure of the quality of $\Psi$.
We refer to the difference between the DMC and HF energies as the DMC
correlation energy, $E_{\rm HF}-E_{\rm DMC}[\Psi]$.
The fraction of the DMC correlation energy retrieved in VMC,
\begin{equation}
f_{\rm DCE}[\Psi] =
  \frac{E_{\rm HF}-E_{\rm VMC}[\Psi]}
       {E_{\rm HF}-E_{\rm DMC}[\Psi]} \;,
\end{equation}
measures the quality of the Jastrow factor, since a perfect Jastrow
factor would make the VMC and DMC energies coincide
\cite{footnote3}.
We define the fraction of the remaining DMC correlation
energy recovered by a wave function $\Psi_2$ with respect to another
$\Psi_1$ as
\begin{equation}
  \frac{E_{\rm VMC}[\Psi_1]-E_{\rm VMC}[\Psi_2]}
       {E_{\rm VMC}[\Psi_1]-E_{\rm DMC}[\Psi_2]} \;.
\end{equation}

The local energy of an electronic configuration ${\bf R}$
is defined as $E_L({\bf R}) = \Psi_{\rm T}^{-1}({\bf R})
{\hat H}({\bf R}) \Psi_{\rm T}({\bf R})$, where ${\hat H}({\bf R})$
is the Hamiltonian operator.
The variance of the local energies encountered in a VMC calculation,
which we shall refer to as the VMC variance, tends to its lower bound
of zero as $\Psi_{\rm T}$ tends to an eigenstate of the Hamiltonian,
and is thus a measure of the overall quality of the trial wave
function.

\subsection{Homogeneous electron gases}
\label{subsec:heg}

\subsubsection{One-dimensional homogeneous electron gas}
\label{subsubsec:1dheg}

We have studied a 1D HEG of density parameter $r_s=5$~a.u.\
consisting of 19 electrons subject to periodic boundary conditions
using a single Slater determinant of plane-wave orbitals.
The ground-state energy of an infinitely thin 1D HEG in which
electrons interact by the
full Coulomb potential is independent of the magnetic state, and hence
we have chosen all the electrons to have the same spin.
This system is unusual in that the nodal surface of the trial function
is exact, and therefore DMC gives the exact ground-state energy, which
we have estimated to be $-0.2040834(3)$ a.u.\ per electron.
Excellent results were reported for this system in
Refs.~\cite{drummond_van_2007,lee_ground-state_2011}
using wave functions with e--e backflow transformations
\cite{kwon_effects_1993,lopez_rios_inhomogeneous_2006}
which preserve the (exact) nodal surface of the Slater determinant.

We have investigated the improvement in VMC results when various terms
are added to an e--e Jastrow factor $J=N_{2,0}+C_{2,0}$, both with and
without backflow transformations.
In the absence of backflow, we find that including $N_{3,0}$,
$C_{3,0}$, or $N_{3,0}+C_{3,0}$ improves the VMC energy, while the
subsequent addition of $C_{4,0}$ yields no further gain.
VMC gives an almost exact energy with backflow and $J=N_{2,0}$,
and therefore no further reduction is possible by including more
Jastrow terms.
However, the addition of $N_{3,0}+C_{3,0}$ reduces the VMC variance
by a factor of five, giving a variance that is an order of magnitude
smaller than that reported in Ref.~\cite{drummond_van_2007} for
a similar calculation.

\subsubsection{Two-dimensional homogeneous electron gas}
\label{subsubsec:2dheg}

We have studied a paramagnetic 2D HEG with 42 electrons per
simulation cell at $r_s=35$~a.u., which lies
close to the Wigner crystallization density predicted by
Drummond and Needs \cite{drummond_phase_2009}.
Kwon {\it et al}.\ \cite{kwon_effects_1993} found that three-electron
correlations are important at low densities, and that the effect of
a three-electron Jastrow factor on the VMC energy is comparable to
that of backflow.
This makes low densities appealing for testing higher-rank Jastrow
terms.

\begin{figure}[h]
  \begin{center}
    \includegraphics[width=0.40\textwidth]{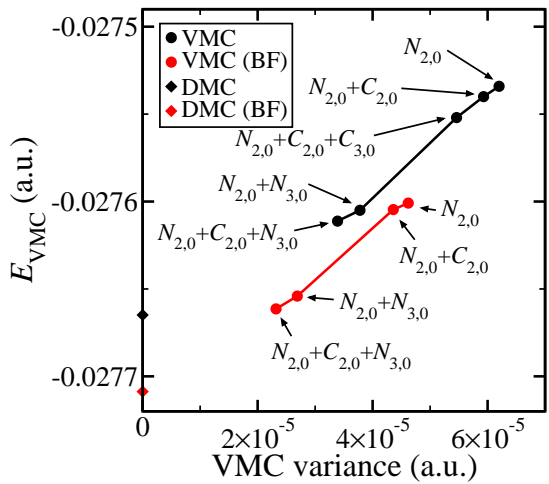}
    \caption{(Color online)
             VMC energies $E_{\rm VMC}$ against the VMC variance for
             the 2D HEG at $r_s=35$~a.u.\ using different Jastrow
             factors, along with the DMC energies for reference.
             The error bars are smaller than the size of the symbols,
             and ``(BF)'' indicates the use of backflow.
             \label{fig:HEG2D_E_vs_V}}
  \end{center}
\end{figure}

The VMC energy and variance obtained using different Jastrow factors
with and without backflow is plotted in Fig.~\ref{fig:HEG2D_E_vs_V}.
The addition of an $N_{3,0}$ term to $J=N_{2,0}$ recovers
81\% of the remaining DMC correlation energy without backflow
and 49\% with backflow.
The $C_{2,0}$ term further reduces both the VMC energy and variance.
The use of a $C_{3,0}$ term recovers 10\% of the remaining DMC
correlation energy when added to $J=N_{2,0}+C_{2,0}$, but it
was not used further since the lack of a cutoff function makes
calculations with $C_{3,0}$ too costly for the little benefit it
provides.

We have also performed DMC calculations using two different Jastrow
factors in the presence of backflow in order to quantify the
indirect effect of the quality of the Jastrow factor on the quality
of the nodes of the wave function.
We obtain a DMC energy of $-0.0277072(1)$~a.u.\ per electron for
$J=N_{2,0}$, and a lower energy of $-0.0277087(1)$~a.u.\ per electron
for $J=N_{2,0}+N_{3,0}+C_{2,0}$.
This supports the idea that a better Jastrow factor allows the
backflow transformation to shift its focus from the ``bulk'' of the
wave function to its nodes, thus improving the DMC energy.

\subsection{Be, B, and O atoms}
\label{subsec:atoms}

While excellent descriptions of these atoms can be obtained within
VMC and DMC using multideterminant wave functions with backflow
correlations \cite{seth_quantum_2011,brown_energies_2007}, we have used
single-determinant wave functions since we are only interested in the
effects of the Jastrow factor.
We have studied the ground states of the Be, B, and O atoms,
corresponding to $^{1}$S, $^{2}$P, and $^{3}$P electronic
configurations, respectively.
The \textsc{atsp2k} code \cite{froese_fischer_mchf_2007} was used
to generate numerical single-electron HF orbitals tabulated
on a radial grid.
We have investigated the use of Jastrow factors with up to four-body
terms, but we have not used backflow for these systems.
The energies of Chakravorty
\textit{et al}.\ \cite{chakravorty_ground-state_1993} have been used
as ``exact'' reference values.

We obtain lower single-determinant VMC energies for the Be, B, and O
atoms with $J=F_{2,0}+F_{1,1}+F_{2,1}$ than reported in
Refs.~\cite{brown_energies_2007,toulouse_full_2008}.
We obtain further small improvements in the VMC energies by including
either $F_{3,0}$ or $F_{3,1}$ Jastrow terms, but their combination,
$F_{3,0}+F_{3,1}$, is not found to be advantageous over using the
terms individually.
This indicates that $F_{3,0}$ and $F_{3,1}$, the latter of which
provides a slightly lower VMC energy than the former, have nearly
the same effect in atoms.
These three-electron terms should be particularly useful in describing
correlations involving electrons in the atomic core region.
We expect $F_{3,1}$ to be more useful than $F_{3,0}$ in molecules
and solids because it should be able to adapt to the different length
scales in these systems, whereas $F_{3,0}$ offers a homogeneous
description of three-electron correlations.
We have investigated the effect of adding a $F_{4,1}$ term in Be and
O, but it does not reduce the VMC energy or variance when added to
$J=F_{2,0}+F_{1,1}+F_{2,1}+F_{3,1}$.

Our best VMC energies of $-14.6522(1)$~a.u.\ (Be),
$-24.6309(2)$~a.u.\ (B), and $-75.0381(3)$~a.u.\ (O), correspond
to fractions of the DMC correlation energy of $94.0(1)\%$, $91.8(1)\%$,
and $94.6(1)\%$, respectively.
The VMC energies are compared with the best single-determinant
nonbackflow VMC values we could find in the literature in
Table~\ref{tab:best_vmc}.

\subsection{BeH, N$_2$, H$_2$O, and H$_2$ molecules}

The BeH, N$_2$, H$_2$O, and H$_2$ molecules are strongly inhomogeneous
and anisotropic systems.
We have used basis sets of moderate quality for the single-electron
orbitals of BeH and N$_2$ in order to investigate the extent to which
the Jastrow factor can compensate for the deficiencies of the basis
sets, especially via one-electron terms $N_{1,m}$.
For H$_2$O and H$_2$ we have used very good basis sets.
We have also tested anisotropic Jastrow factors in N$_2$, and
a van der Waals-like Jastrow factor for H$_2$.

\subsubsection{BeH molecule}
\label{subsec:beh_mol}

We have studied the all-electron BeH molecule in the $^{2}\Sigma^+$
ground state configuration at a bond length of
$2.535$~a.u.\ \cite{feller_survey_2008}.
We have used a single-determinant wave function containing Slater-type
orbitals generated with the \textsc{adf}
package \cite{te_velde_chemistry_2001}, with which we obtain a
reference DMC energy of $-15.24603(4)$~a.u.

The addition of $N_{1,2}$ to $J=N_{2,0}+N_{1,1}+N_{2,1}$ recovers
$11\%$ of the remaining DMC correlation energy.
We find no significant gain from adding either an $N_{2,2}$ term
or an $N_{3,1}$ term to $J=N_{2,0}+N_{1,1}+N_{2,1}+N_{1,2}$.

\subsubsection{N$_2$ molecule}
\label{subsec:n2_mol}

We have studied the $^1\Sigma_g^+$ ground state of the N$_2$ molecule
at the experimental bond length of 2.074~a.u.\ \cite{feller_survey_2008}
HF orbitals were generated in a Slater-type basis using the ADF
package \cite{te_velde_chemistry_2001}.
Our VMC results for different Jastrow factors are given in
Table~\ref{tab:n2_mol} along with relevant reference energies.

\begin{table*}
  \caption{VMC energies ($E$) and variances ($V$) for the N$_2$
           molecule using different Jastrow factors, including
           explicitly anisotropic terms.
           We have used a bond length of $r_{\rm NN} =
           2.074~{\rm a.u}$ \cite{feller_survey_2008}.
           \label{tab:n2_mol}}
\begin{threeparttable}
\begin{tabular}{lrr@{\@}lr@{\@}lr@{\@}lr@{\@}lr@{\@}l}
\hline
\hline
  &
  $N_{p}$ &
  \multicolumn{2}{c}{$E$ (a.u.)} &
  \multicolumn{2}{c}{$V$ (a.u.)} &
  \multicolumn{2}{c}{$f_{\rm CE}$ (\%)} &
  \multicolumn{2}{c}{$f_{\rm DCE}$ (\%)} \\
\hline
 HF limit from Ref.~\cite{oneill_benchmark_2005} &
      &$ -108$&$.9929    $&$ $&$         $&$  0$&$       $&$  0$&$      $
\\[4pt]
 $N_{2,0}$ &
   18 &$ -109$&$.102(1)  $&$ 5$&$.275(4) $&$ 19$&$.9(2)  $&$ 21$&$.3(2) $\\
 $N_{2,0}$+$N_{1,1}$ &
   27 &$ -109$&$.3739(6) $&$ 3$&$.681(3) $&$ 69$&$.4(1)  $&$ 74$&$.3(2) $\\
 $N_{2,0}$+$N_{1,1}$+$N_{1,2}$ &
   49 &$ -109$&$.3796(6) $&$ 3$&$.595(2) $&$ 70$&$.4(1)  $&$ 75$&$.4(2) $\\
 $N_{2,0}$+$N_{1,1}$+$N_{2,1}$ &
   80 &$ -109$&$.4441(4) $&$ 1$&$.667(2) $&$ 82$&$.16(7) $&$ 87$&$.9(1) $\\
 $N_{2,0}$+$N_{1,1}$+$N_{2,1}$+$N_{1,2}$ &
  102 &$ -109$&$.4644(4) $&$ 1$&$.149(2) $&$ 85$&$.85(7) $&$ 91$&$.9(1) $\\
 $N_{2,0}$+$N_{1,1}$+$N_{2,1}$+$N_{1,2}$+$N_{2,2}$ &
  219 &$ -109$&$.4697(4) $&$ 1$&$.088(3) $&$ 86$&$.82(7) $&$ 92$&$.9(1) $\\
 $N_{2,0}$+$N_{1,1}$+$N_{2,1}$+$N_{1,2}$+$N_{2,2}$+$N_{3,0}$ &
  260 &$ -109$&$.4702(3) $&$ 1$&$.083(2) $&$ 86$&$.91(5) $&$ 93$&$.0(1) $
\\[4pt]
 $N_{2,0}$+$N_{1,1}$+$A_{1,1}^z$ &
   36 &$ -109$&$.3770(6) $&$ 3$&$.670(2) $&$ 69$&$.9(1)  $&$ 74$&$.9(2) $\\
 $N_{2,0}$+$N_{1,1}$+$N_{2,1}$+$A_{1,1}^z$ &
   89 &$ -109$&$.4660(3) $&$ 1$&$.116(2) $&$ 86$&$.14(5) $&$ 92$&$.2(1) $\\
 $N_{2,0}$+$N_{1,1}$+$N_{2,1}$+$A_{1,1}^z$+$A_{1,1}^{z^2}$ &
   97 &$ -109$&$.4669(3) $&$ 1$&$.073(2) $&$ 86$&$.31(5) $&$ 92$&$.4(1) $\\
 $N_{2,0}$+$N_{1,1}$+$N_{2,1}$+$A_{1,1}^z$+$A_{2,1}^{z}$ &
  142 &$ -109$&$.4707(3) $&$ 1$&$.072(2) $&$ 87$&$.00(5) $&$ 93$&$.1(1) $\\
 $N_{2,0}$+$N_{1,1}$+$N_{2,1}$+$A_{1,1}^z$+$A_{1,1}^{z^2}$+$A_{2,1}^{z}$+$
  A_{2,1}^{z^2}$ &
  191 &$ -109$&$.4714(3) $&$ 1$&$.036(4) $&$ 87$&$.13(5) $&$ 93$&$.3(1) $
\\[4pt]
 SD VMC from Ref.~\cite{toulouse_full_2008}\tnote{a} &
      &$ -109$&$.4520(5) $&$  $&$        $&$ 83$&$.59(9) $&$ 89$&$.5(2) $\\
 SD DMC &
      &$ -109$&$.5060(7) $&$  $&$        $&$ 93$&$.4(1)  $&$100$&$.0(2) $\\
 Exact from Ref.~\cite{oneill_benchmark_2005} &
      &$ -109$&$.5421    $&$  $&$        $&$100$&$       $&$   $&$      $\\
 \hline
 \hline
\end{tabular}
 \begin{tablenotes}\footnotesize
  \item[a]{For $r_{\rm NN}=2.075~{\rm a.u}$.
           We do not expect that this small difference in bond length
           will affect the comparison between energies significantly.}
 \end{tablenotes}
\end{threeparttable}
\end{table*}

Adding an $N_{1,2}$ term to $J=N_{2,0}+N_{1,1}+N_{2,1}$ factor
recovers $33\%$ of the remaining DMC correlation energy
and leads to a significant reduction in the VMC variance.
The subsequent addition of $N_{2,2}$ provides a reduction
in the VMC energy of $13\%$ of the remaining DMC correlation
energy.  We have tested adding $N_{3,0}$, $N_{3,1}$, and $N_{4,0}$
terms to $J=N_{2,0}+N_{1,1}+N_{2,1}+N_{2,2}$, but neither of
these yield any improvements in the VMC energy.

The anisotropy of this system is expected to be captured by terms
containing e--n functions that treat the bond as a special direction.
We have aligned the $z$-axis of our reference frame along the N--N
bond in our calculations, and $A_{1,1}^z$ is then the simplest
explicitly anisotropic term that reflects the geometry of the system.
The $A_{1,1}^x$ and $A_{1,1}^y$ terms must be zero by symmetry and we
have therefore not used them.
There are five spherical harmonics with $l=2$, which are
respectively proportional to $xy$, $xz$, $yz$, $x^2-y^2$, and
$-x^2-y^2+2z^2$.
In our calculations we find that only the last one of these, which we
refer to as $z^2$, has a significant effect on the VMC energy.

\begin{figure*}
  \begin{center}
    \includegraphics[width=0.60\textwidth]{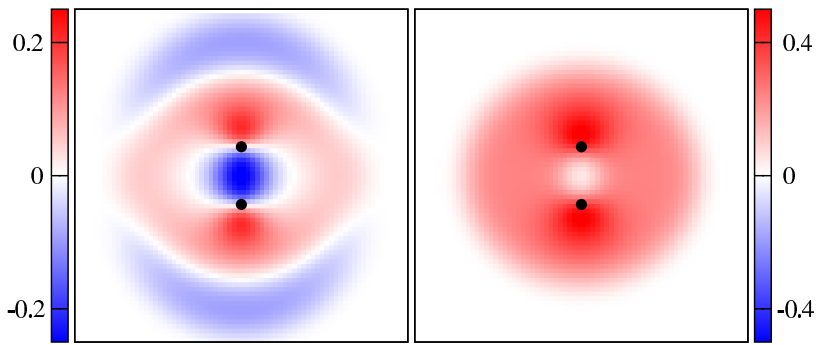}
    \caption{(Color online)
             Plots of the $A_{1,1}^z$ term (left) and $N_{1,2}$ term
             (right) of different Jastrow factors for N$_2$ as a
             function of the position of an electron in a plane
             containing the nuclei (black circles).
             \label{fig:n2_a11z_n12}}
  \end{center}
\end{figure*}

The VMC energy with $J=N_{2,0}+N_{1,1}+N_{2,1}+A_{1,1}^z$ is within
statistical uncertainty of that with $J=N_{2,0}+N_{1,1}+N_{2,1}+
N_{1,2}$, but the former Jastrow factor contains about a third fewer
parameters than the latter.
The combination of the $N_{1,2}$ and $A_{1,1}^z$ terms into
$J=N_{2,0}+N_{1,1}+N_{2,1}+N_{1,2}+A_{1,1}^z$ does not improve the
VMC energy compared with the other two Jastrow factors.
These results suggest that the terms $N_{1,2}$ and $A_{1,1}^z$
play similar roles in the wave function, which we find reasonable
since $N_{1,2}$, although constructed from isotropic basis functions,
contains the right variables to capture the symmetry of the molecule
in much the same way as $A_{1,1}^z$ does.
We have plotted the $A_{1,1}^z$ term for $J=N_{2,0}+N_{1,1}+N_{2,1}+
A_{1,1}^z$ and the $N_{1,2}$ term for $J=N_{2,0}+N_{1,1}+N_{2,1}+
N_{1,2}$ in Fig.~\ref{fig:n2_a11z_n12}, where the similarity between
the terms can be seen.
The value of the $N_{1,2}$ term is roughly the same as that of
$A_{1,1}^z$ offset by a positive amount, and this shift is likely to
be compensated for by the other Jastrow factor terms.
Both terms increase the value of the wave function in the outer
region of the molecule with respect to that in the bond region.

We have added different combinations of anisotropic terms to
$J=N_{2,0}+N_{1,1}+N_{2,1}$.
The e--e--n $A_{2,1}^z$ term retrieves less correlation energy than
the e--n $A_{1,1}^z$ term.
Combining terms with spherical harmonics of $l=1$ and $l=2$ improves
the VMC energy significantly with respect to using $l=1$ only.
The anisotropic Jastrow factor $J=N_{2,0}+N_{1,1}+N_{2,1}+
A_{1,1}^z+A_{1,1}^{z^2}+A_{2,1}^z+A_{2,1}^{z^2}$, which contains up to
e--e--n correlations and has 191 optimizable parameters, recovers
$93.3(1)\%$ of the DMC correlation energy, which is greater than the
$93.0(1)\%$ retrieved by our best isotropic Jastrow factor
$J=N_{2,0}+N_{1,1}+N_{2,1}+N_{1,2}+N_{2,2}+N_{3,0}$, which
includes more costly e--e--n--n and e--e--e correlations and contains
260 optimizable parameters.
We conclude that anisotropic functions are an important tool
in the construction of compact Jastrow factors for strongly
anisotropic systems.

Toulouse and Umrigar obtained $90\%$ of the DMC correlation
energy with a single-determinant wave function in
Ref.~\cite{toulouse_full_2008}, and with our best Jastrow
factor we retrieve $93\%$ of the DMC correlation energy.
We have also optimized a single-determinant backflow wave function
with our best Jastrow factor and we obtain a VMC energy of
$-109.4820(6)$ a.u.\ ($89\%$ of the correlation energy),
which is of similar accuracy to the multideterminant VMC energy of
$-109.4851(3)$ a.u.\ ($89.6\%$ of the correlation energy)
obtained by Toulouse and Umrigar.

\subsubsection{H$_2$O molecule}
\label{subsec:h2o_mol}

Single-particle spin-unrestricted HF orbitals for the
$^1 A_1$ ground state of H$_2$O were generated using the
\textsc{crystal} Gaussian basis set code \cite{dovesi_crystal_2005}.
The basis set for O contains 14 $s$-, 9 $p$-, and 4 $d$-orbitals, and
that for H contains 8 $s$-, 4 $p$-, and 3 $d$-orbitals.
Electron-nucleus cusps have been added using the scheme of Ma
\textit{et al.}\ \cite{ma_scheme_2005}.  We have simulated a water
molecule with a bond length of $r_{\rm OH} = 1.8088$~a.u.\ and a bond
angle of $\angle_{\rm HOH} = 104.52^\circ$
\cite{gurtubay_dissociation_2007}.

Adding an $N_{1,2}$ term to $J=N_{2,0}+N_{1,1}+N_{2,1}$ gives only a
very small improvement for H$_2$O, compared with the more substantial
improvements obtained with this e--n--n term for BeH and N$_2$.
The $N_{1,2}$ term acts as a correction to the single-electron orbitals,
and we believe that it is unimportant in this case because we have used
very accurate HF orbitals, whereas the single-electron orbitals used for
BeH and N$_2$ are considerably less accurate.
We find additional small improvements to the energy of H$_2$O from
adding $N_{3,0}$ and $N_{3,1}$ terms to $J=N_{2,0}+N_{1,1}+N_{2,1}$.

Clark {\it et al}.\ obtained $92\%$ of the DMC correlation
energy with a single-determinant wave function in
Ref.~\cite{clark_computing_2011}, and with our best Jastrow
factor we recover $95.5\%$ of the DMC correlation energy.

\subsubsection{H$_2$ molecule}
\label{subsec:h2_mol}

The energy of the first triplet excited state ($^3\Sigma_u^+$) of
H$_2$ has a very shallow minimum corresponding to a large bond length
of nearly $8$~a.u.
Although the exchange interaction falls exponentially with increasing
internuclear separation, Kolos and Wolniewicz found that it
contributed significantly to the energy even at the large distance of
$10$~a.u.\ \cite{kolos_potentialenergy_1965}
The strong interplay between the attractive dispersion forces and
the repulsive exchange interaction requires that both be accounted
for to afford an accurate description of the triplet state.
This makes the system appealing for studying the construction of
four-body Jastrow factor terms to describe van der Waals-like
interactions.

We generated numerical HF orbitals tabulated on an elliptical grid
using the \textsc{2dhf} package \cite{laaksonen_fully_1986}.
HF theory predicts no binding for the triplet state at any
separation, and therefore any binding that occurs in VMC can be
attributed to the Jastrow factor.
We have studied the H$_2$ molecule in its triplet spin state at the
internuclear distance of $7.8358$~a.u.
This separation and the corresponding energy of $-1.0000208957$~a.u.\
were found by fitting a quadratic function to the data of Staszewska
and Wolniewicz \cite{staszewska_transition_1999}.

Previous QMC calculations on H$_2$ at different
interatomic distances have used Jastrow factors with up to
four-body correlations where the cusp conditions were not
enforced \cite{alexander_ground_2004,per_anisotropic_2009},
instead relying on the variance minimization method to find
parameter values that approximately satisfy the cusp conditions.
This was found to be advantageous for this system because the
additional variational freedom yielded a better description
in VMC than when the cusp conditions were obeyed
exactly \cite{per_personal_2012}.
The violation of the cusp conditions is potentially catastrophic
in DMC calculations, but previous studies have restricted the
use of such terms to VMC\@.

\begin{figure}[h]
  \begin{center}
    \includegraphics[width=0.40\textwidth]{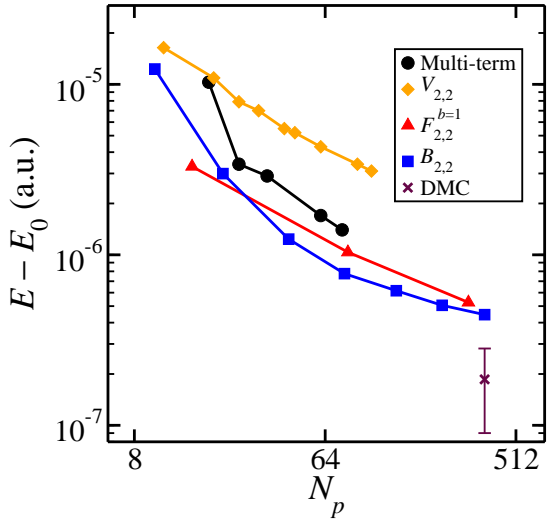}
    \caption{(Color online)
             Difference between the VMC and exact energy against
             the number of wave function parameters for the
             H$_2$ triplet ground-state using different Jastrow
             factors.
             Only the ``multiterm'' Jastrow factor enforces the
             cusp conditions.
             The error bars are smaller than the size of the symbol
             where not shown.
             All of the energies in this plot are lower than
             $-1$ a.u.\ and therefore the wave functions
             predict binding.
             \label{fig:h2_e_vs_np}}
  \end{center}
\end{figure}

For H$_2$ we have optimized Jastrow factors consisting of the single
e--e--n--n terms $V_{2,2}$, $F_{2,2}^{b=1}$, and $B_{2,2}$ (see
Table~\ref{tab:notation}) at several expansion orders, where no
constraints are enforced at e--e or e--n coalescence points.
We have used variance minimization for these Jastrow factors as we
found that it produces better results than energy minimization.
We have also optimized Jastrow factors consisting of different sums
of terms which satisfy the cusp conditions using energy minimization.
The results are shown in Fig.~\ref{fig:h2_e_vs_np}.

We have performed the DMC calculations using our best $B_{2,2}$
Jastrow factor and obtain a reference DMC energy of
$-1.0000207(1)$~a.u.
We have not encountered any statistical problems in the DMC
calculations with this cusp-violating wave function.
Such issues can occur when the local energy presents a negative
divergence in a region of configuration space with a significant
probability of being sampled.
We have verified that our wave function causes a negative divergence
in the local energy when an electron coalesces with a nucleus,
a point at which the wave function is likely to be relatively large.
We therefore conclude that the region of influence of this divergence
is sufficiently small that statistical problems do not arise in
practice.

The $F_{2,2}^{b=1}$ and $B_{2,2}$ terms only differ in that the latter
uses Boys-Handy-style indexing, which yields slightly lower VMC
energies than standard indexing in most cases for a fixed number of
parameters.
Our best $F_{2,2}^{b=1}$ and $B_{2,2}$ Jastrow factors retrieve
$99\%$ of the DMC correlation energy in VMC\@.

The $V_{2,2}$ term is designed to describe van der Waals correlations,
and contains e--e functions which introduce other correlations.
Our best $V_{2,2}$ term recovers $92\%$ of the DMC correlation
energy, offering a good description of the system without reaching the
accuracy of the more generic $F_{2,2}^{b=1}$ and $B_{2,2}$ terms.

A $V_{2,2}$ term without e--e functions consists of contributions
proportional to ${\bf r}_{iI} \cdot {\bf r}_{jJ}$, where the
prefactors depend explicitly on $r_{iI}$ and $r_{jJ}$, and
implicitly on $r_{IJ}$, and this functional form is that of a
dipole-dipole interaction potential.
Our best such $V_{2,2}$ term retrieves $69\%$ of the DMC correlation
energy, which amounts to $0.0000262(3)$ a.u., and we regard this
as a measure of the pure van der Waals correlation energy of this
system.

The multiterm Jastrow factors contain the usual $N_{2,0}$, $N_{1,1}$,
$N_{1,2}$, and $N_{2,1}$ terms, and for each combination of these we
have added a $V_{2,2}$ term without e--e functions obeying the cusp
conditions to study its effect.
$J=N_{2,0}$ retrieves $44\%$ of the DMC correlation energy, and
adding the $V_{2,2}$ term retrieves $85\%$ of the remaining DMC
correlation energy.
The effectiveness of $V_{2,2}$ progressively drops as more terms are
added, and it retrieves $43\%$ of the remaining DMC correlation energy
when added to $J=N_{2,0}+N_{1,1}+N_{2,1}+N_{1,2}$.
In all cases, $V_{2,2}$ is found to lower the VMC energy by a
larger amount than any of the $N_{n,m}$ terms.

Our best multiterm cusp-enforcing Jastrow factor retrieves
$97\%$ of the DMC correlation energy with 77 wave-function
parameters, comparable with the $98\%$ retrieved with the
cusp-violating $F_{2,2}^{b=1}$ and $B_{2,2}$ terms with a similar
number of parameters.
For larger systems where van der Waals interactions are important,
we expect the violation of cusp conditions to cause statistical
problems, and the $V_{2,2}$ term would become an effective way of
improving the description of the system in a multiterm Jastrow
factor.

\subsection{Discussion of molecular results}
\label{sec:discussion}

\begin{figure}[h]
  \begin{center}
    \includegraphics[width=0.40\textwidth]{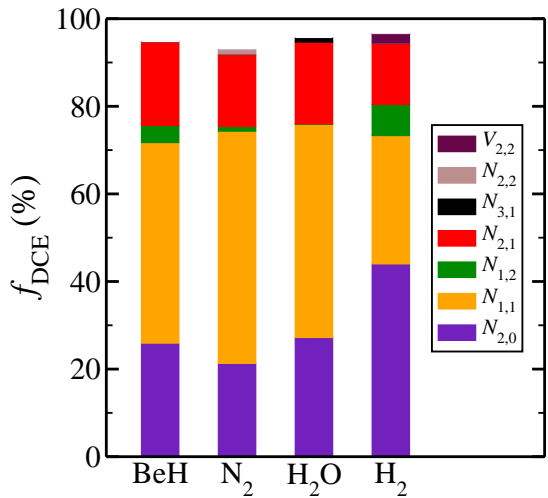}
    \caption{(Color online)
             Fraction of the DMC correlation energy retrieved by
             different Jastrow factor terms, added in the specific
             order depicted in the graph (starting from the bottom),
             for the molecules studied in this paper.
             \label{fig:mol_bar_chart}}
  \end{center}
\end{figure}

In Fig.~\ref{fig:mol_bar_chart} we have plotted the fraction
of the DMC correlation energy retrieved by different Jastrow factor
terms for the BeH, N$_2$, H$_2$O, and H$_2$.
Our purpose is to visualize the importance of different terms in
different system, and to this end we do not include anisotropic
or cusp-violating terms, and we only consider the addition of
terms in a specific order.

The $N_{2,0}$ term represents the simplest description of electronic
correlations and typically retrieves 20--25\% of the DMC correlation
energy.
This e--e term greatly distorts the charge density of the HF wave
function, and the $N_{1,1}$ term repairs this, retrieving an
additional 45--50\% of the DMC correlation energy.
In the case of the more diffuse H$_2$ molecule the $N_{2,0}$ and
$N_{1,1}$ terms have a different relative importance, but
$J=N_{2,0}+N_{1,1}$ retrieves 70--75\% of the DMC correlation
energy in the four molecules.

Like $N_{1,1}$, $N_{1,2}$ acts as a correction to the single-electron
orbitals.
This term provides no significant benefit in H$_2$O, where we have
used high-quality orbitals, but it recovers $7\%$ of the DMC
correlation energy in H$_2$.

The effect of $N_{1,2}$ in N$_2$ is noteworthy in that the energy
reduction obtained by adding this term to $J=N_{2,0}+N_{1,1}$ is about
a factor of four times smaller than when added to the more accurate
$J=N_{2,0}+N_{1,1}+N_{2,1}$.
One would expect a term to retrieve more correlation energy when added
to a smaller Jastrow factor, and this is the case for $N_{1,2}$ in the
other molecules.
We think that the distortion in the charge density caused by
$N_{2,1}$ in N$_2$ is such that the single-electron correction
effected by $N_{1,2}$ becomes more useful in its presence.

The $N_{2,1}$ term added to $J=N_{2,0}+N_{1,1}+N_{1,2}$ captures an
additional 15--20\% of the DMC correlation energy.
Higher-order terms added to $J=N_{2,0}+N_{1,1}+N_{2,1}+N_{1,2}$
yield significant gains in relative terms,
with e--e--n--n terms retrieving 13\% and 43\% of the remaining DMC
correlation energy remaining for N$_2$ and H$_2$, respectively, and
the e--e--e--n term recovering 17\% of the remaining DMC correlation
energy.

\subsection{Summary of results}

Table \ref{tab:best_vmc} gives a comparison of the best
single-determinant nonbackflow VMC energies we have found in the
literature with those obtained in this work.

\begin{table*}
  \caption{Best single-determinant nonbackflow VMC energies (a.u.)
           found in the literature and those from this work, along
           with single-determinant DMC and exact energies for
           reference.
           \label{tab:best_vmc}}
\begin{threeparttable}
\begin{tabular}{lr@{\@}lr@{\@}lr@{\@}lr@{\@}l}
\hline
\hline
 System &
 \multicolumn{2}{c}{This work} &
 \multicolumn{2}{c}{Literature} &
 \multicolumn{2}{c}{DMC} &
 \multicolumn{2}{c}{Exact} \\
\hline
1D HEG ($r_s$=5, $N$=19)
  &$  -0$&$.2040833(2) $&$    $&$                  $
  &$  -0$&$.2040834(3) $&$  -0$&$.2040834(3)$\\
2D HEG ($r_s$=35, $N$=42)
  &$  -0$&$.0276112(6) $&$    $&$                  $
  &$  -0$&$.0277087(1) $&$    $&$           $\\
Be
  &$ -14$&$.6522(1)    $&$ -14$&$.64972(5)\tnote{a}$
  &$ -14$&$.65717(4)   $&$ -14$&$.66736     $\\
B
  &$ -24$&$.6309(2)    $&$ -24$&$.62936(5)\tnote{a}$
  &$ -24$&$.64002(6)   $&$ -24$&$.65391     $\\
O
  &$ -75$&$.0381(3)    $&$ -75$&$.0352(1) \tnote{a}$
  &$ -75$&$.0511(1)    $&$ -75$&$.0673      $\\
BeH
  &$ -15$&$.2412(3)    $&$ -15$&$.228(1)  \tnote{b}$
  &$ -15$&$.24603(4)   $&$ -15$&$.2482      $\\
N$_2$
  &$-109$&$.4714(3)    $&$-109$&$.4520(5) \tnote{c}$
  &$-109$&$.5060(7)    $&$-109$&$.5421      $\\
H$_2$O
  &$ -76$&$.4068(2)    $&$ -76$&$.3938(4) \tnote{d}$
  &$ -76$&$.4226(1)    $&$ -76$&$.438       $\\
H$_2$ (${}^3\Sigma_u^+$)
  &$  -1$&$.00002045(3)$&$    $&$                  $
  &$  -1$&$.0000207(1) $&$  -1$&$.0000208957$\\
 \hline
 \hline
\end{tabular}
 \begin{tablenotes}
  \item[a]{Ref.~\cite{toulouse_full_2008}.}
  \item[b]{Ref.~\cite{nemec_benchmark_2010}.}
  \item[c]{Ref.~\cite{toulouse_full_2008} (using a slightly
           different bond length).}
  \item[d]{Ref.~\cite{clark_computing_2011}.}
 \end{tablenotes}
\end{threeparttable}
\end{table*}

\section{Conclusions}
\label{sec:conclusions}

We have described a generalized Jastrow factor allowing terms that
explicitly correlate the motions of $n$ electrons with $m$ static
nuclei.
These terms can be parameterized using various basis sets, including
terms that involve dot products of interparticle position vectors.
We have also introduced anisotropic cutoff functions.
The formalism may be applied to systems with particle types and
external potentials other than electrons and Coulomb potentials.

Optimization of the wave function is one of the most human- and
computer-time consuming tasks in performing QMC
calculations.
We have performed term-by-term optimizations to understand how
different terms in the Jastrow factor contribute to the electronic
description of a system, and we hope that our analysis will serve
as a guideline for constructing Jastrow factors for other
systems.

We have tested these terms on HEGs, atoms, and
molecules.
The variational freedom from the higher-order terms generally improves
the quality of the wave function.
We have only considered single-determinant wave functions in this
study, although our Jastrow factor can of course be used with other
wave function forms.

We have demonstrated the construction and application of an e--e--n--n
Jastrow factor term designed to describe van der Waals interactions
between atoms.
This term retrieves a large fraction of the van der Waals correlation
energy in tests on the triplet state of H$_2$ at the proton
separation that minimizes the total energy of the system.

We have found evidence for the importance of three-electron Jastrow
terms in the low-density 1D and 2D HEGs.
Improving the Jastrow factor for single-determinant backflow wave
functions also leads to small improvements in the DMC energy of the 2D
HEG\@.
This demonstrates the indirect effect that improving the Jastrow
factor can have on improving the nodal surface, as reported in
Ref.~\cite{kwon_effects_1993}.

We have made efforts to obtain accurate single-determinant
VMC energies for most of the systems studied, but for BeH and N$_2$ we
deliberately used inferior one-electron basis sets to see whether
we could compensate for this with one-electron Jastrow terms.  We find
that this goal can be achieved by including an $N_{1,2}$ Jastrow term
or anisotropic e--n terms, along with the usual $N_{1,1}$ term.

\begin{acknowledgments}
We thank John Trail for producing the HF wave function for H$_2$.
The calculations were performed on the Cambridge High Performance
Computing Service facility.
\end{acknowledgments}

\appendix
\section{Implementation}
\label{app:implementation}

In this section we describe our design choices in implementing
our Jastrow factor in the \textsc{casino} code
\cite{needs_continuum_2010}.
The implementation principles are modularity and extensibility,
embracing the flexibility that this Jastrow factor has by design
rather than impeding it by focusing too strictly on performance.

\subsection{Basis functions}

The most important design requirement for modularity is that
basis sets be dealt with separately rather than included in the
Jastrow factor code.
Basis functions, along with their derivatives when required, are
evaluated and stored for later use.
For any given term with expansion orders $p$ and $q$ there are
$pN(N-1)/2$ e--e and $qNM$ e--n functions to evaluate and
store.
Single-electron updates involve recalculating $p(N-1)$ and $qM$ of
these functions.
Note that the number of basis functions that must be evaluated is
independent of the term ranks $n$ and $m$.
Furthermore, it is possible to allow different terms to share
basis functions that do not contain optimizable parameters so that,
e.g., the natural powers involved in computing $N_{2,0}$
and $N_{1,1}$ can be re-used for $N_{2,1}$.

A number of properties of the basis functions are required to
construct the Jastrow factor.
Equation (\ref{eq:lambda_symmetry_invariant_signature}) requires knowledge
of whether basis functions are symmetric or antisymmetric,
and their value, first radial derivative, and angular
dependency at the origin are required by Eqs.\ (\ref{eq:ee_cusp}),
(\ref{eq:ee_noaniso}), (\ref{eq:ee_nocusp}), (\ref{eq:en_cusp}),
(\ref{eq:en_noaniso}), and (\ref{eq:en_nocusp}).
The one-contribution, no-contribution, and equal-product constraints
in these equations require a table indexing distinct products of two
basis functions at any value of their arguments.
We implement interfaces that make these properties avaliable so that
basis functions can be treated as abstract objects in the
construction of the Jastrow factor, which makes implementing new
basis sets straightforward.

Cutoff functions are dealt with as additional basis sets with an
expansion order of one, and we store information identifying the
cutoff functions that are strictly zero, which we use to speed up
evaluation of the Jastrow factor.

\subsection{Evaluation of the Jastrow factor}

For the evaluation of an arbitrary-rank Jastrow factor term it
is necessary to use efficient procedures that iterate from a
given set of electronic and nuclear indices $\ivec$ and $\Ivec$ to the
next in a specific order; explicit loops over scalar integer indices
are not an option in static code since the loop depth is variable,
and the memory usage of precomputing all possible $\ivec$ and $\Ivec$
scales badly with system size for high ranks $n$ and $m$.
These procedures should take into account which cutoff functions are
zero so that particle sets that do not contribute to the Jastrow
factor are skipped.
Efficient handling of localized Jastrow factor terms is important
because it allows the cost of evaluating a term of any rank to
scale linearly with system size if the cutoff lengths are held
fixed.
We implement a scheme where we construct a list of the electrons
that are ``connected'' to each electron and each nucleus via nonzero
cutoff functions.
For the electronic indices, the value of $i_1$ is iterated between $1$
and $N$, then the value of $i_2$ is iterated over the values in the
list associated with electron $i_1$ that are greater than $i_1$, then
the value of $i_3$ is iterated over the values in the intersection
between the lists associated with $i_1$ and $i_2$ that are greater
than $i_2$, and so on.
The procedure that iterates over nuclear indices selects sets of
nuclei whose connected-electron lists have nonzero intersections.
We iterate over $\Ivec$ in the outermost loop so that we can feed the
intersection of all e--n lists into the $\ivec$ iterator as an initial
list for index $i_1$.

The signature $\{{\tilde\Pmat}(\ivec),{\tilde\Smat}(\ivec,\Ivec)\}$
of each group of particles is computed inside the electronic and
nuclear loops, identifying the linear parameter channel associated
with the group of particles.
We then loop over linear parameters in the channel, computing the
products of the relevant basis functions which have already been
precomputed.
In terms without indexing constraints, consecutive linear parameters
tend to have very similar expansion indices, i.e., they
multiply most of the same basis functions.
In order to save multiplications, it is convenient to buffer partial
products so that, e.g., if only the last two of six expansion
indices change from one parameter to the next, we can recover the
product of the first four functions from the previous index set and
save three of the five multiplications required to combine the six
basis functions.

In typical QMC calculations individual electron
moves, rather than full configuration moves, are proposed, which
requires computing an acceptance probability involving the ratio of
the trial wave function at the proposed and original positions.
To calculate this efficiently one needs to compute the part of the
Jastrow factor which depends on the position of a single particle $i$,
ignoring the contributions not involving $i$.
In our implementation we evaluate this one-electron
Jastrow factor using Eq.\ (\ref{eq:gjastrow_sum_of_core}), where we
fix $i_1=i$ and iterate over the rest of $\ivec$.
The main difference from the evaluation of the full Jastrow factor is
that $\ivec$ is not sorted, and the permutation required to sort
$\ivec$, which amounts to inserting $i$ at the correct position in
$\ivec$, needs to be taken into account in the presence of
antisymmetric e--e basis functions, since sign changes will be
required by Eq.\ (\ref{eq:lambda_symmetry_invariant_signature}).
The evaluation of gradients and Laplacians of the Jastrow factor term
can be easily accommodated in the one-electron Jastrow evaluation
code.

For performance reasons it is advisable to implement versions of
the evaluation procedure for fixed ranks, with fixed-depth
loops that can be optimized by compilers. We implement three optimized
versions: one for e--e terms, one for e--n terms, and one for
e--e--n terms.  Other terms are handled by three generic procedures:
one for terms without e--e functions, one for terms without e--n
functions, and one for terms with both e--e and e--n functions.

\section{Term construction example}
\label{app:example}

Let us consider the $N_{3,0}$ term used for the 2D HEG in
Section~\ref{subsubsec:2dheg}.
This is a system with $N=42$ electrons, half of each spin, and the
$N_{3,0}$ term corresponds to $n=3$, $m=0$, expansion order
$p=4$, and basis functions
\begin{equation}
\Phi_\nu^P({\bf r}) =
  r^{\nu-1}
  \left( \frac{L_P-r}{L_P} \right) \;.
\end{equation}
The spin-pair dependency matrix $\Pmat$ is of size $N\times N$, but in
practice we specify a reduced $2\times 2$ version of this matrix
where each row (column) corresponds to different spins.
For this system this matrix is
\begin{equation}
\Pmat =
\left(
  \begin{matrix}
    1 & 2 \\
    2 & 1
  \end{matrix}
\right) \;,
\end{equation}
where we assign parallel-spin electron pairs a spin-pair dependency
index of 1, and antiparallel-spin electron pairs an index of 2.
The $\Pmat$ matrix can be regarded as an intrinsic property of the
system, but in some cases additional symmetries can be imposed in
order to reduce the number of parameters in the Jastrow factor term;
for example we would achieve this by setting all elements of $\Pmat$
to 1 in the 2D HEG, choosing to ignore the distinction between
parallel- and antiparallel-spin electron pairs.

The elements in $\Pmat$ determine pairwise properties and objects.
For example, there are as many sets of nonlinear parameters in a
Jastrow factor term as different values in $\Pmat$; in this case,
there are two cutoff lengths, $L_1$ for parallel-spin electron
pairs and $L_2$ for antiparallel-spin electron pairs.

Four distinct types of three-electron groups can be formed:
three up-spin electrons ($\uparrow\uparrow\uparrow$),
two up-spins and one down-spin ($\uparrow\uparrow\downarrow$),
one up-spin and two down-spins ($\uparrow\downarrow\downarrow$),
and three down-spin electrons ($\downarrow\downarrow\downarrow$).
The spin-pair dependency matrices for these four groups are
\begin{equation}
\Pmat(\uparrow\uparrow\uparrow) =
\left(
  \begin{matrix}
    0 & 1 & 1 \\
    1 & 0 & 1 \\
    1 & 1 & 0
  \end{matrix}
\right) \;\;;\;
\Pmat(\downarrow\downarrow\downarrow) =
\left(
  \begin{matrix}
    0 & 1 & 1 \\
    1 & 0 & 1 \\
    1 & 1 & 0
  \end{matrix}
\right) \;\;
\end{equation}
\begin{equation}
\Pmat(\uparrow\uparrow\downarrow) =
\left(
  \begin{matrix}
    0 & 1 & 2 \\
    1 & 0 & 2 \\
    2 & 2 & 0
  \end{matrix}
\right) \;\;;\;
\Pmat(\uparrow\downarrow\downarrow) =
\left(
  \begin{matrix}
    0 & 2 & 2 \\
    2 & 0 & 1 \\
    2 & 1 & 0
  \end{matrix}
\right) \;.
\end{equation}
The matrices $\Pmat(\uparrow\uparrow\uparrow)$ and
$\Pmat(\downarrow\downarrow\downarrow)$ are identical, and therefore
correspond to a linear parameter channel which is used for groups of
three electrons with parallel spins.
The matrix $\Pmat(\uparrow\downarrow\downarrow)$ can be transformed
into $\Pmat(\uparrow\uparrow\downarrow)$ via, e.g., the permutation
matrix
\begin{equation}
\Umat =
\left(
  \begin{matrix}
    0 & 0 & 1 \\
    0 & 1 & 0 \\
    1 & 0 & 0
  \end{matrix}
\right) \;,
\end{equation}
and therefore both matrices correspond to a second linear parameter
channel which is used for groups of three electrons with mixed spins.
The signature of the first channel, which we refer to as the
$\uparrow\uparrow\uparrow$ channel, is
$\Pmat(\uparrow\uparrow\uparrow)$, and the signature of the
$\uparrow\uparrow\downarrow$ channel is
$\Pmat(\uparrow\uparrow\downarrow)$.
Both of these matrices are considered sorted by our matrix-sorting
algorithm.

The symmetry constraints for the parameters in each of the two
channels depend on the above matrices.
$\Pmat(\uparrow\uparrow\uparrow)$ is invariant with respect to the
application of any permutation, and for the
$\uparrow\uparrow\uparrow$ channel
Eq.~(\ref{eq:lambda_symmetry_invariant_signature}) equates any
two parameters with the same indices in a different order.
The signature of the $\uparrow\uparrow\downarrow$ channel is only
invariant with respect to one nontrivial permutation,
\begin{equation}
\Umat =
  \left(
    \begin{matrix}
      0 & 1 & 0 \\
      1 & 0 & 0 \\
      0 & 0 & 1
    \end{matrix}
  \right) \;,
\end{equation}
and therefore the symmetry constraints for this channel are
\begin{equation}
\lambda_{
  \left(
    \begin{matrix}
      0        & \nu_{ij} & \nu_{ik} \\
      \nu_{ij} & 0        & \nu_{jk} \\
      \nu_{ik} & \nu_{jk} & 0
    \end{matrix}
  \right)
}^{
  \left(
    \begin{matrix}
      0 & 1 & 2 \\
      1 & 0 & 2 \\
      2 & 2 & 0
    \end{matrix}
  \right)
} = \lambda_{
  \left(
    \begin{matrix}
      0        & \nu_{ij} & \nu_{jk} \\
      \nu_{ij} & 0        & \nu_{ik} \\
      \nu_{jk} & \nu_{ik} & 0
    \end{matrix}
  \right)
}^{
  \left(
    \begin{matrix}
      0 & 1 & 2 \\
      1 & 0 & 2 \\
      2 & 2 & 0
    \end{matrix}
  \right)
} \;.
\end{equation}

We do not allow expansion indices to be zero in the $N_{3,0}$ term,
and the presence of cutoff functions prevents the ``one contribution''
constraint of Eq.~(\ref{eq:ee_cusp}) from being satisfied.
It is therefore not possible to impose the Kato cusp conditions on
the $N_{3,0}$ term.
This is not a problem in practice because we use this term in
conjunction with $N_{2,0}$.

Since the basis functions are isotropic, the only constraint
applicable to this term at coalescence points is
Eq.~(\ref{eq:ee_nocusp}) for $l=m=0$.
The derivative of the radial projection of the basis function is
\begin{equation}
\left[
  \frac {\partial {\mathcal P}_{0,0}\left[
         \Phi_\nu^P({\bf r})\right]}
        {\partial r}
\right]_{r=0} =
\delta_{\nu 2} - \delta_{\nu 1} \frac{C}{L_P} \;,
\end{equation}
and the constraint equation for index $\nu_{ij}$ is
\begin{equation}
\sum_{\nu_{ij}\nu_{ik}\nu_{jk}}^{\rm e.p.}
\left[
\lambda_{
  \left(
    \begin{matrix}
      0        & \nu_{ij} & \nu_{ik} \\
      \nu_{ij} & 0        & \nu_{jk} \\
      \nu_{ik} & \nu_{jk} & 0
    \end{matrix}
  \right)
}^{
  \left(
    \begin{matrix}
      0      & P_{ij} & P_{ik} \\
      P_{ij} & 0      & P_{jk} \\
      P_{ik} & P_{jk} & 0
    \end{matrix}
  \right)
}
\delta_{\nu_{ij} 2} - \delta_{\nu_{ij} 1} \frac{C}{L_{P_{ij}}}
\right]
= 0 \;.
\end{equation}
Any two products of pairs of natural powers is equal if the sum of the
exponents in each of them is equal.
Therefore the ``equal-product'' constraint for the $N_{3,0}$ term is
\begin{equation}
\sum_{\nu_{ik}\nu_{jk}}^{\rm e.p.} {} =
\sum_{\nu_{ik}=1}^{\min(l-1,p)} \delta_{\nu_{jk},l-\nu_{ik}} \;,
\end{equation}
where $l$ ranges from 2 to $2p$, and the constraint equation for
index $\nu_{ij}$ thus reduces to
\begin{widetext}
\begin{equation}
\sum_{\nu_{ik}=1}^{\min(l-1,p)}
\left[
\lambda_{
  \left(
    \begin{matrix}
      0        & 2          & \nu_{ik} \\
      2        & 0          & l-\nu_{ik} \\
      \nu_{ik} & l-\nu_{ik} & 0
    \end{matrix}
  \right)
}^{
  \left(
    \begin{matrix}
      0      & P_{ij} & P_{ik} \\
      P_{ij} & 0      & P_{jk} \\
      P_{ik} & P_{jk} & 0
    \end{matrix}
  \right)
} - \lambda_{
  \left(
    \begin{matrix}
      0        & 1          & \nu_{ik} \\
      1        & 0          & l-\nu_{ik} \\
      \nu_{ik} & l-\nu_{ik} & 0
    \end{matrix}
  \right)
}^{
  \left(
    \begin{matrix}
      0      & P_{ij} & P_{ik} \\
      P_{ij} & 0      & P_{jk} \\
      P_{ik} & P_{jk} & 0
    \end{matrix}
  \right)
} \frac{C}{L_{P_{ij}}}
\right]
= 0 \;.
\end{equation}
\end{widetext}
The constraints for the three expansion indices are equal by symmetry
in the $\uparrow\uparrow\uparrow$ channel, and there are two sets of
constraints in the $\uparrow\uparrow\downarrow$ channel.



\begin{thebibliography}{99}
\bibitem{footnote1} 
  For example, $\sum_{i\neq j}^{N} f_{i,j}$ contains $N(N-1)$ terms,
  but if $f_{i,j}=f_{j,i}$, this operation can be simplified to
  $\sum_{i<j}^N f_{i,j}$, which contains $N(N-1)/2$ terms.
  In general, the sum of an object which is symmetric with respect
  to the interchange of any of its $n$ indices, where these indices
  take nonrepeated values, can be rewritten as a sum over sorted sets
  of indices involving $n!$ times fewer terms than the original sum.
\bibitem{footnote2} 
  Given a symmetric $n \times n$ matrix $\Smat$ and a
  rectangular $n\times m$ matrix $\Pmat$, this algorithm finds the
  permutation of these matrices, $\Ptilde=\Umat\MM\Pmat\MM\UmatT$ and
  $\Stilde=\Umat\MM\Smat\MM\RmatT$, that correspond to the lowest
  collational order according to a predefined sorting criterion.
  The specifics of such criterion are unimportant; the only
  requirement is that for any $n \times n$ permutation matrix $\Amat$
  and $m \times m$ permutation matrix $\Bmat$, the application of the
  algorithm to the pair of matrices $\Amat\MM\Pmat\MM\AmatT$ and
  $\Amat\MM\Smat\MM\BmatT$ also returns $\Ptilde$ and $\Stilde$.
\bibitem{footnote3} 
  In general, attaining the DMC limit in VMC would require
  nonanalyticities in the Jastrow factor.  Nonetheless this theoretical
  limit is useful for assessing the performance of a Jastrow factor.
\end{thebibliography}
\end{document}